\documentclass[twocolumn,english,aps,reprint, superscriptaddress,showpacs,longbibliography,showkeys,pra]{revtex4-2}
\usepackage{amsmath,amssymb,bbm,dsfont, mathrsfs,bm,braket,color,graphicx,comment} 
\usepackage[table, dvipsnames]{xcolor}
\usepackage{tabularray}
\usepackage[colorlinks,citecolor=blue,urlcolor=blue,linkcolor = blue]{hyperref}
\usepackage[mathscr]{euscript}
\usepackage{enumitem}
\usepackage{dsfont}
\usepackage{amsthm}
\usepackage{nicematrix}
\usepackage{blkarray}
\usepackage{marginnote}
\usepackage{multirow}
\usepackage{algorithm}
\usepackage{algpseudocode}
\usepackage{bbold}
\usepackage{booktabs}
\usepackage{xspace}
\usepackage{tikz}
\usetikzlibrary{arrows.meta, positioning}

\usepackage{ulem} 

\newcommand{\cluster}{\textsc{Cluster}\xspace}
\newcommand{\ebadi}{\textsc{GreedyMIS}\xspace}

\begin{document}

\title{Error-Corrected Quantum Optimization via Rydberg Atom Array Embeddings}

\author{Andrew Byun}
\email[These authors contributed equally.\\Corresponding authors: \\]{woojeong.byun@uibk.ac.at, berend.klaver@uibk.ac.at}
\affiliation{Institut f\"{u}r Theoretische Physik, Universit\"{a}t Innsbruck, A-6020 Innsbruck, Austria}

\author{Berend Klaver}
\email[These authors contributed equally.\\Corresponding authors: \\]{woojeong.byun@uibk.ac.at, berend.klaver@uibk.ac.at}
\affiliation{Institut f\"{u}r Theoretische Physik, Universit\"{a}t Innsbruck, A-6020 Innsbruck, Austria}
\affiliation{Parity Quantum Computing GmbH, A-6020 Innsbruck, Austria}

\author{Arinjoy De}
\affiliation{QuEra Computing Inc., Boston, MA 02139, USA}

\author{Pedro Ildefonso}
\affiliation{Institut f\"{u}r Theoretische Physik, Universit\"{a}t Innsbruck, A-6020 Innsbruck, Austria}
\affiliation{Parity Quantum Computing GmbH, A-6020 Innsbruck, Austria}

\author{Davit Khachatryan}
\affiliation{Parity Quantum Computing GmbH, A-6020 Innsbruck, Austria}

\author{Alexei Bylinskii}
\affiliation{QuEra Computing Inc., Boston, MA 02139, USA}

\author{Michael Schuler}
\affiliation{Parity Quantum Computing GmbH, A-6020 Innsbruck, Austria}

\author{Wolfgang Lechner}
\affiliation{Institut f\"{u}r Theoretische Physik, Universit\"{a}t Innsbruck, A-6020 Innsbruck, Austria}
\affiliation{Parity Quantum Computing GmbH, A-6020 Innsbruck, Austria}
\affiliation{Parity Quantum Computing Germany GmbH, 20095 Hamburg, Germany}
\affiliation{Parity Quantum Computing France SAS, 75016 Paris, France}

\begin{abstract}
Rydberg atom quantum optimizers naturally encode the maximum independent set problem on geometric graphs. However, to solve general combinatorial optimization problems on more complex topologies, an \textit{embedding} is required to overcome the limited-range native interactions. 
Such embeddings use \textit{gadgets}, specific arrangements of atoms, which inevitably introduce ancillary atoms.
Here, we show that this apparent overhead instead provides an intrinsic resource for error correction.
Specifically, we demonstrate that the Parity Architecture-based Rydberg embedding naturally forms a classical error-correcting code, a principle that extends to other Rydberg blockade-based embeddings.
Within \textit{superatom} substructures, exact Rydberg-excitation-number constraints provide local error syndromes at no additional hardware cost and give rise to a strictly nonlinear code.
To correct bit-flip errors, we introduce the precisely tailored \cluster decoder, which exploits these local constraints and the underlying gadget structure.
Numerically, this approach achieves an error threshold of $p_{\text{th}}\gtrsim26 \%$ across different layout geometries.
Experimentally, on the QuEra Aquila QPU, the \cluster decoder recovers exact optimization solutions with an average success rate of $\gtrsim8\%$ for systems of up to 107 atoms, even when bare hardware and standard greedy post-processing find none.
A Rydberg atom array embedding is therefore not merely a hardware mapping of an optimization problem, but a co-designed architecture for quantum optimization in which embedding, computation, and error correction emerge from the same physical structure.
\end{abstract}

\date{\today}
\maketitle

\section{Introduction}

\begin{figure*}[tb]
    \centering
\includegraphics[width=1.0\linewidth]{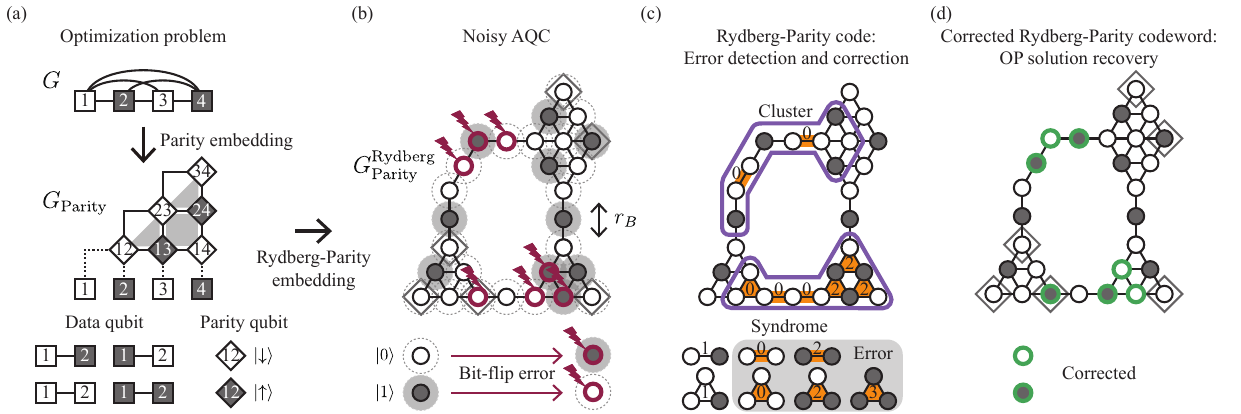}
    \caption{Overview of error-corrected quantum optimization via Rydberg atom array embeddings.
    (a) An optimization problem (OP) on a logical graph $G$ (here, a QUBO instance) is first mapped to the parity encoding $G_{\text{Parity}}$. Squares denote data qubits and diamonds denote parity qubits, with empty (filled) markers indicating the qubit state $\ket{\downarrow}$ ($\ket{\uparrow}$). 
    Gray triangles and squares represent three- and four-body parity constraints, respectively. 
    (b) The parity-encoded problem is embedded into a Rydberg-blockade-based atomic layout $G_{\text{Parity}}^{\text{Rydberg}}$, yielding the Rydberg-Parity embedding. Adiabatic quantum computing (AQC) is then performed on this layout. Empty (filled) circles correspond to physical atoms in the ground state $\ket{0}$ (Rydberg state $\ket{1}$). 
    Due to hardware and diabatic noise, the raw AQC readout suffers from bit-flip errors (highlighted in red). 
    (c) The Rydberg-Parity embedding intrinsically acts as a classical error-correcting code: Rydberg-Parity code. Because the Rydberg blockade energetically enforces exactly one excitation per \textit{superatom} for valid configurations, an error syndrome is extracted by locally counting Rydberg excitations: superatoms containing zero or multiple excitations are flagged as errors (bottom).
    Our proposed \cluster decoder groups the flagged superatoms into local clusters (purple outlines) and efficiently corrects the corresponding errors using the underlying gadget structure.
    (d) Finally, the \cluster decoder outputs the corrected physical bitstring, from which the solution to the original optimization problem on $G$ can be directly extracted.
}
    \label{fig:1}
\end{figure*}

Quantum optimization~\cite{Kadowaki1998, Farhi2000, Farhi2001, Das2008, Farhi2014, Peruzzo2014, Albash2018, Cerezo2021, Bharti2022, Abbas2024}, i.e., solving optimization problems (OPs) using quantum devices, represents one of the most promising current and near-term applications of quantum technology.
Within this domain, adiabatic quantum computing (AQC) provides a hardware-friendly alternative to circuit-based approaches for noisy intermediate-scale quantum (NISQ) devices~\cite{Preskill2018, Bharti2022}. 
Among current platforms, Rydberg atom arrays~\cite{Browaeys2020, Semeghini2021, Ebadi2021, Scholl2021} are particularly attractive because of their scalability and the flexible spatial arrangement of qubits, making them a leading platform for near-term quantum optimization~\cite{Wurtz2024MIS}.
In particular, Rydberg atom arrays natively encode the maximum (weight) independent set (M(W)IS) problem on unit-disk (UD) graphs through the intrinsic \textit{Rydberg blockade} mechanism~\cite{Pichler2018, Pichler2018NP}. 
Over the past decade, significant experimental progress has been made, including solving large-scale UD-MIS problems on arrays of several hundred qubits~\cite{Ebadi2022}, unit-ball MIS problems in three-dimensional configurations~\cite{Song2021, Kim2022}, and UD-MWIS instances enabled by local detuning control~\cite{DeOliveira2025, DeOliveira2026}.
However, native quantum optimization on Rydberg atom arrays is largely restricted to UD-M(W)IS problems on planar geometric graphs. 
To solve arbitrary combinatorial OPs beyond UD-MWIS, an additional \textit{embedding} step is required~\cite{Choi2008, Lechner2015}. 
This can be achieved using Rydberg \textit{gadgets}: atom arrangements that implement specific constraints and allow connections between distant variables. 
Prominent embedding strategies include the Parity Architecture-based embedding~\cite{Lanthaler2023, Lanthaler2026}, crossing lattices~\cite{Nguyen2023, DeOliveira2025, Bombieri2025}, and 3D quantum wires~\cite{Kim2022, Byun2022}. 
By appropriately assembling Rydberg gadgets into larger atomic structures, a wide variety of hard combinatorial OPs, including satisfiability (SAT)~\cite{Jeong2023, Park2024}, integer factorization~\cite{Nguyen2023, Park2024}, many-body spin models~\cite{Stastny2023, Zeng2025, Maier2025}, and general QUBO or HUBO problems~\cite{Lanthaler2023, Lanthaler2026, Nguyen2023, Byun2024, Byun2024HUBO}, can be systematically mapped onto the native Rydberg atom hardware.

A fundamental consequence of Rydberg gadget embeddings is the introduction of auxiliary atoms, which increases the physical system size and the overall exposure to physical errors. 
Because full active quantum error correction remains out of reach for near-term analog devices, classical post-processing and architectural error mitigation are essential for improving their performance.
Current mitigation strategies for quantum optimization include quantum annealing correction~\cite{Jordan2006, Ganti2014} for superconducting platforms~\cite{Pudenz2014, Pudenz2015, Vinci2015, Vinci2016, MunozBauza2025}, as well as broader theoretical frameworks that exploit constraint-based redundancy~\cite{Lechner2015, Rocchetto2016, Pastawski2016, Weidinger2023}. 
For neutral atoms specifically, standard approaches rely on heuristic greedy algorithms (such as \ebadi~\cite{Ebadi2022, Wurtz2024}), generic post-selection~\cite{Kim2022, Byun2022}, or state-deselection techniques~\cite{DeCorrec2025}.
However, these existing mitigation methods generally ignore the intricate spatial structure of the Rydberg gadget embeddings or suffer from exponential scaling.
Specifically, the interplay between the spatial geometry of the gadgets and the strict physical constraints of the Rydberg blockade has not yet been systematically exploited to correct errors.

In this work, we introduce a new perspective on Rydberg-blockade-based embeddings: the redundancy built into the physical layout itself forms a \textit{classical error-correcting code}~\cite{Hamming1950}.
Focusing primarily on the Rydberg-Parity embedding~\cite{Lanthaler2023, Lanthaler2026}, we show that the embedding naturally groups atoms into \textit{superatoms}, each constrained by the Rydberg blockade to host exactly one Rydberg excitation in valid logical states.
By simply counting local excitations, we can therefore extract an \textit{error syndrome}, converting the bare embedding into the \textit{Rydberg-Parity code}.
Because these constraints are counting-based, the Rydberg-Parity code is strictly \textit{nonlinear}, rendering standard linear-code decoders inapplicable.
To address this, we introduce the \cluster decoder, a tailored, scalable algorithm that resolves errors locally by explicitly leveraging the Rydberg gadget structure. 
We demonstrate that this co-design of hardware embedding and software decoding exhibits a finite error-correction threshold of $p_{\text{th}} \gtrsim 26\%$ under a simple bit-flip noise model. 
Furthermore, using experimental data from the QuEra Aquila QPU~\cite{Wurtz2023Aquila}, we show that the \cluster decoder significantly enhances quantum optimization success rates for layouts of up to 107 atoms. 

Figure~\ref{fig:1} summarizes our full workflow for error-corrected quantum optimization with Rydberg gadget embeddings. 
An optimization problem is first mapped to the physical Rydberg-Parity embedding on an array of atoms via an intermediate parity mapping  [Fig.~\ref{fig:1}(a, b)]. 
Then, noisy AQC is performed on this Rydberg-Parity layout [Fig.~\ref{fig:1}(b)], typically yielding a noisy bitstring. Interpreting the atomic layout as a Rydberg-Parity code, we extract the error syndrome, apply the \cluster decoder to correct errors, and recover a valid bitstring [Fig.~\ref{fig:1}(c)]. 
Finally, the solution of the original OP is directly obtained from the corrected bitstring [Fig.~\ref{fig:1}(d)].

This work is structured as follows: In Sec.~\ref{sec:embedding_review}, we review the Rydberg-Parity embedding for combinatorial optimization on neutral atom arrays. 
In Sec.~\ref{sec:error_correction}, we establish the core theoretical contribution of this work: formalizing the embedding as a classical, nonlinear error-correcting code, introducing the \cluster decoder, and numerically demonstrating its error threshold. 
In Sec.~\ref{Sec:QPU}, we benchmark the performance of this scheme on experimental QPU data. 
Finally, we present our conclusions and outlook in Sec.~\ref{sec:conclusion}. 
Further details of our approach and the extension to crossing lattice and quantum wire embeddings are provided in the Appendix.

\section{Review: Rydberg gadget embeddings}
\label{sec:embedding_review}

In this section, we review the embedding of combinatorial OPs into Rydberg atom arrays using Rydberg gadgets, providing the theoretical foundation for our work. 
We first introduce the necessary background of the considered quantum hardware platform: a Rydberg-blockade-based UD-MWIS solver. 
Following this, we detail the Rydberg-Parity embedding~\cite{Lanthaler2023}, which unlocks the embedding of arbitrary combinatorial OPs, and focus on its two primary building blocks (LINK and 3BODY gadgets) and the amalgamation mechanism used to assemble them into complex layouts.

\subsection{Rydberg-blockade-based UD-MWIS solver}
The quantum hardware platform we consider is an optical-tweezer-arranged Rydberg atom system~\cite{Browaeys2020}; the fully filled, arbitrarily shaped atomic array is tuned by optical tweezers~\cite{Kim16, Barredo16, Endres16}, laser-cooled beforehand, and thus effectively frozen on the timescale of the individual experiments.
%
Rydberg atom qubits are encoded in the atomic ground state $\ket{0}$ and a Rydberg state $\ket{1}$. Two atoms simultaneously in the Rydberg state interact via the van der Waals potential $U_{ij}=C_6/|\mathbf{r}_i-\mathbf{r}_j|^6$, where $C_6>0$ is the interaction coefficient and $\mathbf{r}_i$ and $\mathbf{r}_j$ denote the positions of the atoms. 
The full Hamiltonian of a Rydberg atom array is written as (in units where $\hbar=1$),

\begin{subequations}
    \begin{eqnarray}
        \hat{H}_{\text{Ryd}} &=& \hat{H}_{\text{mix}}+\hat{H}_{\text{cost}}, \\
        \hat{H}_{\text{mix}} &=& \sum_i \frac{\Omega}{2} \hat{X}_i ,\label{eq:mix_hamiltonian}\\
        \hat{H}_{\text{cost}} &=& -\sum_i \Delta_i \hat{n}_i + \sum_{i<j} U_{ij} \hat{n}_i\hat{n}_j, \label{eq:MWIS_hamiltionian}
    \end{eqnarray}
\end{subequations}
where $\hat{X}_i = \ket{0}_i\bra{1}+\ket{1}_i\bra{0}$ and $\hat{n}_i=\ket{1}_i\bra{1}$ are the Pauli-X and Rydberg number operators for atom $i$, respectively. $\Omega$ denotes the global Rabi frequency, and $\Delta_i$ denotes the local detuning of the atom at site $i$.

The Rydberg blockade energetically penalizes simultaneous excitation of atoms $i$ and $j$ to the Rydberg state if they are within the blockade radius ${r_B=(C_6/\sqrt{\Delta^2+\Omega^2})^{1/6}}$, i.e., $|\mathbf{r}_i-\mathbf{r}_j| \le r_B$. 
To find the ground-state(s) of $\hat{H}_{\text{cost}}$ via AQC in a Rydberg-atom system, one typically first initializes all atoms in state $\ket{0}$, sets the Rabi frequency to $\Omega = 0$ and the detunings to negative values, $\Delta_i < 0$, so that the initialized state is the ground state of $\hat{H}_\text{Ryd}$. 
Then, the Rabi frequency is slowly turned on ($\Omega > 0$) while the detunings are swept to their final target values $\Delta_i > 0$. 
While approaching the target detunings, the Rabi frequency is finally turned off again. 

In the limit
\begin{equation}
\Omega \rightarrow 0, \quad 0 < \Delta_k < U_{ij} \ ,
\label{eq:mwis_condition}
\end{equation}
the many-body ground state of $\hat{H}_{\text{Ryd}}=\hat{H}_{\text{cost}}$ encodes the UD-MWIS of the corresponding UD graph ${G=(V,E; w)}$~\cite{Pichler2018}, where each vertex $v_i \in V$ is represented by an atom, each edge $(v_i,v_j) \in E$ connects a blockaded pair of atoms, and each vertex weight $w_i$ is proportional to the site-dependent detuning, $w_i \propto \Delta_i$~\footnote{Rydberg atom arrays can also be arranged in 3D geometries~\cite{Barredo2018, Song2021, Kim2022, Kusano2025} and in that setup encode unit-ball MWIS problems. For simplicity of the discussion, we here only consider 2D geometries.}. 
The MWIS problem can be defined as:
\[
\begin{aligned}
&{\text{Maximize}}\sum_{v_i \in V} w_i n_i, \nonumber\\
&\text{with}\quad n_i n_j \neq 1 \ {\text{if}} \: (v_i,v_j) \in E,
\end{aligned}
\]
where $n_i \in \{0,1\}$ are binary variables associated with the vertices $v_i$.
In particular, $n_i=1$ means that $v_i \in V$ is included in the MWIS.
In the Rydberg atom setup, the variables $n_i$ correspond to the eigenvalues of the number operators $\hat{n}_i$.
Note that finding the MWIS of a graph $G$, including the UD-MWIS case, is known to be NP-complete~\cite{Pichler2018, Pichler2018NP}. The UD constraint arises from the distance-dependent van der Waals interaction $U_{ij} \propto |\mathbf{r}_i-\mathbf{r}_j|^{-6}$, which generates blockade only between nearby atoms satisfying the condition in Eq.~\eqref{eq:mwis_condition}. From now on, for simplicity, we assume the infinite-blockade approximation, in which 
\begin{eqnarray}
        U_{ij}\approx
\begin{cases}
\infty & |\mathbf{r}_i-\mathbf{r}_j| \le r_B, \\
0 & |\mathbf{r}_i-\mathbf{r}_j|>r_B.
\end{cases}
\label{eq:PXP}
\end{eqnarray}
Under the condition in Eq.~\eqref{eq:mwis_condition} and the infinite-blockade
approximation in Eq.~\eqref{eq:PXP}, any configuration violating the blockade is excluded from the
low-energy Hilbert space. Consequently, the cost Hamiltonian
$\hat{H}_{\text{cost}}$ effectively reduces to $\hat{H}_{\text{cost}}\rightarrow \hat{H}^{\text{BLC}} $
\begin{equation}
    \hat{H}^{\text{BLC}} = \mathcal{P} \left( - \sum_{v_i \in V} \Delta_i \hat{n}_i \right) \mathcal{P},
    \label{eq:H_BLC}
\end{equation}
where $\mathcal{P}$ denotes the projection into the subspace of Rydberg-blockade satisfying states, i.e., independent set configurations~\cite{Verrensen2021, Stastny2023}.
We will refer to the Hamiltonian $\hat{H}^{\text{BLC}}$ as the \textit{blockaded Hamiltonian} and, for brevity, drop the projectors $\mathcal{P}$ hereafter.

After the adiabatic sweep, a typical Rydberg experiment performs a projective measurement of the Rydberg occupation numbers $n_i \in \{0, 1\}$ of the atomic array. Therefore, each shot returns a \textit{classical bitstring} $\mathbf{n} = (n_1 n_2 \cdots)$ corresponding to the measurement-basis state $\ket{\mathbf{n}}$.
If AQC successfully prepares a ground state of $\hat{H}_\text{cost}$, the corresponding classical bitstring represents a solution to the UD-MWIS problem.

\subsection{Rydberg-Parity embedding}
\label{sec:gadgets}

\begin{figure}[t]
    \centering
    \includegraphics[width=1.0\linewidth]{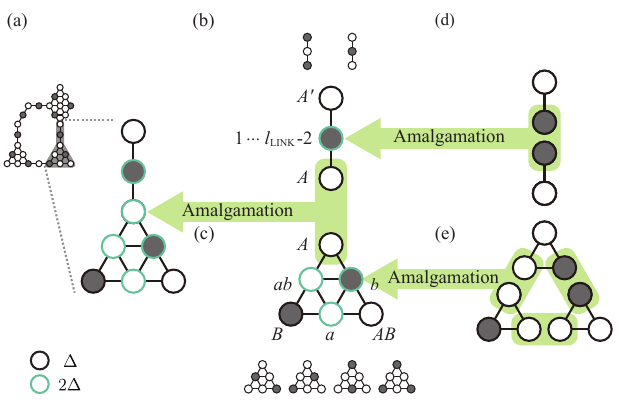}
    \caption{Gadget-based Rydberg-Parity embedding. 
    (a) The physical layout $G_{\text{Parity}}^{\text{Rydberg}}$ is constructed through the \textit{amalgamation} of two fundamental gadgets~\cite{Lanthaler2023}: 
    (b) The LINK gadget, which copies the bit from one end to the other, $n_A = n_{A'}$;
    (c) The 3BODY gadget, which satisfies the parity constraint: $n_A \oplus n_B \oplus n_{AB} = 1$. 
    Amalgamation glues gadgets together at shared atoms (e.g., atom $A$), thereby summing their local detunings $\Delta_i$ (indicated by the circle colors).
    The smaller configurations shown in (b) and (c) enumerate all valid MWIS bitstrings of the corresponding gadgets.
    (d, e) LINK and 3BODY gadgets can themselves be interpreted as an amalgamation of dimer and trimer \textit{superatoms}. The Rydberg blockade strictly enforces an exactly-one Rydberg excitation constraint on each superatom.
    In all panels, empty (filled) circles correspond to atoms in the ground state $\ket{0}$ (Rydberg state $\ket{1}$).
    }
    \label{fig:2}
\end{figure}

To solve problems beyond UD-MWIS, Rydberg atom arrays require gadget-based embeddings~\cite{Kim2022, Lanthaler2023, Nguyen2023, Stastny2023, Byun2024, Byun2024HUBO, Lanthaler2026}.
In this paper, we focus on the \textit{Rydberg-Parity embedding}~\cite{Lanthaler2023, Lanthaler2026}; an overview of the general embedding concept and alternative schemes is provided in App.~\ref{App:embedding}.
Embeddings $G \rightarrow G'$ generally come with a space overhead.
For example, embedding QUBO or MWIS problems on all-to-all connected graphs of size $|V|=N$ within a Rydberg atom array generally requires $|V'|=\mathcal{O}(N^2)$ physical atoms.

The Rydberg-Parity embedding~\cite{Lanthaler2023, Lanthaler2026} is performed in two steps [see Fig.~\ref{fig:1}(a, b)]
\begin{equation}
    G \rightarrow G_{\text{Parity}} \rightarrow G_{\text{Parity}}^{\text{Rydberg}} \ .
    \label{eq:parity_embedding}
\end{equation}
In particular, an unconstrained combinatorial OP (QUBO, HUBO) defined on a graph $G$ is first mapped onto a ground state problem on a different graph $G_{\rm Parity}$ via the parity mapping~\cite{Lechner2015, Ender2023, terHoeven2023}. The resulting problem involves only single-qubit weights and local three- and four-body constraints [see Fig.~\ref{fig:1}(a); further details are provided in App.~\ref{App:Parity}].
Then, using Rydberg gadgets to implement the constraints and required linking patterns, this problem is embedded into a UD-MWIS problem on $G_{\text{Parity}}^{\text{Rydberg}}$, which can be realized with a Rydberg atom array~\cite{Lanthaler2023}.
Through this embedding, the MWIS of $G_{\text{Parity}}^{\text{Rydberg}}$ contains the solution of the original OP on $G$. 

Specifically, the layout of $G_\text{Parity}^\text{Rydberg}$ is determined by the structure of the OP graph $G$ through the intermediate parity-mapped graph $G_\text{Parity}$. The problem coefficients, denoted by $\mathbf{J}=(J_1,J_2,\dots)$, determine the weights of the target OP but do not alter this spatial structure; instead, they are encoded through local detunings applied to specific atoms in $G_{\text{Parity}}^{\text{Rydberg}}$~\cite{Lanthaler2023, Lanthaler2026}.
At $\mathbf{J}=0$, all valid logical states form a degenerate ground-state manifold, while applying local detunings ($\mathbf{J}\neq0$) lifts this degeneracy and selects the ground state(s) solving the target OP.
Because our error correction scheme introduced below relies solely on the geometric structure and local constraints of the embedding, it is independent of the specific problem coefficients $\mathbf{J}$. 
Consequently, we set $\mathbf{J}=0$ throughout the theoretical discussion without loss of generality, while Sec.~\ref{Sec:QPU} experimentally benchmarks instances with $\mathbf{J}\neq0$.

We now focus on the final layout $G_{\text{Parity}}^{\text{Rydberg}}$.
Figure~\ref{fig:2}(a) shows a \textit{gadget}-assembled embedded layout. 
Gadgets are specialized atomic layouts with tailored weight patterns, designed to embed binary relations directly within their MWIS configurations.
In particular, Figs.~\ref{fig:2}(b,c) show the two fundamental gadgets to construct $G_{\text{Parity}}^{\text{Rydberg}}$: LINK and 3BODY gadgets.
By gluing these gadgets together into larger structures, a UD-MWIS graph $G_\text{Parity}^\text{Rydberg}$ can be constructed for any parity-mapped problem $G_\text{Parity}$.
This gluing is done by the \textit{amalgamation} process~\cite{Lanthaler2023, Nguyen2023,Stastny2023}, as illustrated in Fig.~\ref{fig:2}.
In the following, we define these two gadgets and the amalgamation method.

The LINK gadget [Fig.~\ref{fig:2}(b)] is used to copy the bit information from the atom at the beginning of the LINK, $A$, to the atom at the end of the LINK, $A'$, satisfying the constraint $C_{\text{LINK}}$:
\begin{equation}
    C_{\text{LINK}}: n_{A} = n_{A'} \ .
    \label{Const_LINK}
\end{equation}
We refer to the atoms $A$ and $A'$ as \textit{ports}, which encode logical information and can be connected to ports of other gadgets.
A LINK gadget is realized by an odd number of quasi-linearly arranged atoms with weight pattern $\Delta_A = \Delta_{A'} = \Delta$, and $\Delta_i = 2\Delta$ for all interior atoms.
Therefore, the blockaded Hamiltonian for the LINK gadget of length $l_{\text{LINK}}$ (number of atoms) is described by
\begin{equation}
    \hat{H}_{\text{LINK}}^{\text{BLC}} = -\Delta \left(\hat{n}_A + \hat{n}_{A'}\right) - 2 \Delta \sum_{i=1}^{l_{\text{LINK}}-2} \hat{n}_i \ .
    \label{eq:link_hamiltonian}
\end{equation}
The ground states of this Hamiltonian in the blockade-satisfying subspace are the two antiferromagnetically ordered states
\begin{equation}
\label{eq:link_state}
\ket{n_A; n_1\cdots; n_{A'}}=\ket{0;10\cdots 01;0} \text{ and } \ket{1;01\cdots10;1}, 
\end{equation}
which exactly copy information between the two ports, $n_A = n_{A'}$, as desired.

The 3BODY gadget [Fig.~\ref{fig:2}(c)] implements the 3-body parity constraint between three parity qubits labeled $A$, $B$, and $AB$~\cite{Lanthaler2023}:
\begin{equation}
    C_{\text{3BODY}}: n_A \oplus n_B \oplus n_{AB} = 1 \, ,
    \label{Const_3BODY}
\end{equation}
where $\oplus$ denotes bitwise XOR.
Implemented as an atomic gadget, it is composed of three ports corresponding to the parity qubits ($n_A, n_B, n_{AB}$) and three auxiliary atoms ($n_a, n_b, n_{ab}$) with detunings
$\Delta$ and $2\Delta$ respectively, arranged in a triangular pattern as shown in Fig.~\ref{fig:2}(c). 
The blockaded Hamiltonian for the 3BODY gadget is 
\begin{equation}
    \hat{H}_{\text{3BODY}}^{\text{BLC}} = -\Delta \left(\hat{n}_A + \hat{n}_B + \hat{n}_{AB}\right) - 2 \Delta \left( \hat{n}_a + \hat{n}_b + \hat{n}_{ab}\right) \ .
    \label{eq:3body_hamiltonian}
\end{equation}
There are four possible MWIS states, i.e., ground states of $H_{\text{3BODY}}^{\text{BLC}}$ in the constrained subspace: 
\begin{equation}
\label{eq:3body_states}
\begin{aligned}
\ket{n_An_Bn_{AB};n_an_bn_{ab}}
={}&\ket{100;100}, \ket{010;010}, \\
&\ket{001;001}, \text{ and } \ket{111;000}.
\end{aligned}
\end{equation}
As required, they exactly and uniquely fulfill the 3-body parity constraint from Eq.~\eqref{Const_3BODY}.

Amalgamation glues multiple gadgets together along shared logical variables. 
For example, to amalgamate a LINK and a 3BODY gadget, which share the logical variable $n_A$, we glue the two gadgets on the nodes representing $n_A$ and sum up the assigned weights from each gadget [see Fig.~\ref{fig:2}(a)]. Therefore, the amalgamation corresponds to the AND conjunction of the two constraints in Eq.~\eqref{Const_LINK} and Eq.~\eqref{Const_3BODY}:
\begin{eqnarray}
    C_{\text{amal}}=C_{\text{LINK}}\wedge C_{\text{3BODY}}.
    \label{Const_Amal}
\end{eqnarray}
At the level of the blockaded Hamiltonian this conjunction is realized by summing the Hamiltonians of the two gadgets, in which the detunings assigned to the shared atoms add up,
\begin{eqnarray}
\label{eq:amal_hamiltonian}
    \hat{H}^{\text{BLC}}_{\text{amal}} &=& \hat{H}_{\text{LINK}}^{\text{BLC}}+\hat{H}_{\text{3BODY}}^{\text{BLC}} \nonumber \\
    &=&\left(-\Delta \hat{n}_A +\cdots\right) + \left(-\Delta \hat{n}_A+\cdots \right)  \nonumber \\
    &=&-2\Delta \hat{n}_A +\cdots.
    \label{Eq:HamilAmal}
\end{eqnarray}
The ground-state energy of $\hat{H}^{\text{BLC}}_{\text{amal}}$ is bounded from below by the sum of the ground-state energies of the two summands, and the bound is attained if and only if a configuration exists that satisfies both constraints. In that case, the two summands are necessarily minimized simultaneously, so every MWIS of the amalgamated graph restricts exactly to an MWIS on each individual gadget~\cite{Stastny2023}.

This local optimality makes error correction possible: a global solution can be validated subgraph by subgraph without solving the full problem. In the following section, we leverage this property to define error syndromes for Rydberg-gadget embedded structures and demonstrate our error correction procedure.

\section{Error correction for Rydberg gadget embeddings}
\label{sec:error_correction}

The fundamental principle of error correction is to use redundant information to detect and correct errors, giving rise to a \textit{code}.
In the Rydberg-gadget approach, mapping an $N$-variable OP onto an embedded layout of $\mathcal{O}(N^2)$ atoms naturally introduces such redundancy, allowing the spatial embedding itself to function as a \textit{classical error-correcting code}.
The structure underlying this redundancy is an amalgamation of \textit{superatom} subgraphs, within which valid configurations satisfy exact Rydberg-excitation-number constraints.
Violations of these local constraints can be detected simply by counting excitations, directly providing an \textit{error syndrome}.
Crucially, the resulting code is \textit{nonlinear}, rendering standard linear-code decoders inapplicable.
To correct errors, we therefore introduce the specifically tailored \cluster decoder, which exploits the local gadget structure and the nonlinear constraints on superatoms, and dynamically expands the decoding region when a unique correction cannot be identified.

\subsection{Superatom amalgamation}

In Sec.~\ref{sec:gadgets} we have reviewed how Rydberg gadgets can be glued into larger layouts through amalgamation by summing up local detunings of shared nodes.
We now show that the gadgets themselves can be constructed in the same way by amalgamating Rydberg \textit{superatoms}~\cite{Lukin01, Dudin2012, Zeiher2015, Labuhn2016}, i.e., groups of atoms where all pairwise distances fall within the Rydberg blockade radius, thereby restricting the superatom to have at most one Rydberg excitation.
Even more, the condition in Eq.~\eqref{eq:mwis_condition} enforces an \textit{exactly-one} Rydberg excitation condition for superatoms in the low-energy subspace.
We define a \textit{superatom set} $S \subseteq V(G_{\text{Parity}}^{\text{Rydberg}})$ as the set of atoms that belong to the same superatom. Therefore, the exactly-one excitation constraint $C_S$ on a bitstring $\mathbf{n}$ is satisfied if and only if:
\begin{eqnarray}
\label{eq:constraint_superatom}
    C_{S}(\mathbf{n}): \left( \sum_{v_i \in S} n_i = 1 \right).
\end{eqnarray}

Figure~\ref{fig:2}(d) shows that the LINK gadget of length $l_{\text{LINK}}$ is an amalgamation of $l_{\text{LINK}}-1$ dimers, which are minimal superatoms of size two. The corresponding constraint is written as a conjunction of the dimer superatom constraints,
\begin{align}
C_{\text{LINK}}^{\text{Rydberg}} &=  C_S(n_A,n_1) \wedge C_S(n_1,n_2) \wedge \cdots \nonumber \\
 & \wedge C_S(n_{l_{\text{LINK}}-2},n_{A'}) \ ,
\end{align}
and satisfies $C_{\text{LINK}}$ from Eq.~\eqref{Const_LINK}.
To see why, note that each dimer forces its two atoms to alternate, and the amalgamation repeats this pattern. 
Then, since $l_{\text{LINK}}$ is odd, the two ports are guaranteed to have identical bit values, which is the condition $C_{\text{LINK}}$. 
In addition, $C_{\text{LINK}}^{\text{Rydberg}}$ also fixes the auxiliary atoms to have a $\mathbb{Z}_2$ ordering, and reduces to $C_{\text{LINK}}$ on the ports.
From a different perspective, summing the dimer Hamiltonians reproduces 
$\hat{H}_\text{LINK}^\text{BLC}$ [Eq.~\eqref{eq:link_hamiltonian}]:
\begin{eqnarray}
\hat{H}_{\text{LINK}}^{\text{BLC}} &=& \underbrace{-\Delta (\hat{n}_A+\hat{n}_1)}_{C_S(n_A,n_1)} \underbrace{-\Delta (\hat{n}_1+\hat{n}_2)}_{C_S(n_1,n_2)} + \cdots \nonumber \\
& &\underbrace{-\Delta (\hat{n}_{l_{\text {LINK}}-2}+\hat{n}_{A'})}_{C_S(n_{l_{\text{LINK}}-2}, n_{A'})}.
\end{eqnarray}

Analogously, Fig.~\ref{fig:2}(e) shows that the 3BODY gadget is an amalgamation of three trimers (superatoms of size three) with the constraint
\begin{eqnarray}
\label{eq:3body_constraint}
C_{\text{3BODY}}^{\text{Rydberg}}&=& C_S(n_A,n_b,n_{ab})  \wedge C_S(n_B,n_a,n_{ab})  \nonumber \\ 
& & \wedge C_S(n_{AB},n_a,n_b) \ .
\end{eqnarray}
This constraint satisfies $C_{\text{3BODY}}$ in Eq.~\eqref{Const_3BODY} at the ports of the triangle.
In particular, the sum of the trimer Hamiltonians reproduces $\hat{H}_\text{3BODY}^\text{BLC}$ [Eq.~\eqref{eq:3body_hamiltonian}],
\begin{align}
\hat{H}^{\text{BLC}}_{\text{3BODY}}&=\underbrace{-\Delta(\hat{n}_A+\hat{n}_b+\hat{n}_{ab})}_{C_S(n_A,n_b,n_{ab})} \nonumber \\
& 
\underbrace{-\Delta(\hat{n}_B+\hat{n}_a+\hat{n}_{ab})}_{C_S(n_B,n_a,n_{ab})} 
\underbrace{-\Delta(\hat{n}_{AB}+\hat{n}_a+\hat{n}_{b})}_{C_S(n_{AB},n_a,n_{b})}.
\end{align}

In summary, we have shown that the embedding to $G_{\text{Parity}}^{\text {Rydberg}}$ can be regarded as a repeated structure of dimer and trimer superatoms, each fulfilling an exactly-one excitation constraint in their low-energy (logical) subspace. 
Note that crossing lattice gadgets can also be decomposed into superatoms in this manner, whereas quantum wires require a generalized local constraint [see App.~\ref{App:CRS} and App.~\ref{App:Wire}].

\subsection{Superatom syndrome}

By using the exactly-one excitation constraints imposed by the superatoms in Eq.~\eqref{eq:constraint_superatom}, a simple \textit{number-check} provides a local error-detection mechanism for bit-flip errors, analogous to a parity check in classical error-correcting codes. Here, we define the \textit{superatom syndrome} $\sigma$ associated with a superatom set $S$ based on the number check:
\begin{equation}
\label{eq:superatom_syndrome}
\sigma_{S}(\mathbf{n})= \begin{cases}  \mathbb{0}, & \text { if }  C_{S}(\mathbf{n})=\text{True} ,\\ 
\mathbb{1}, & \text { if }  C_{S}(\mathbf{n})= \text{False} .
        \end{cases}
\end{equation}

We assign $\mathbb{0}$ to each satisfied exactly-one constraint and $\mathbb{1}$ for each unsatisfied case, as in standard error syndrome conventions, and use the symbols $\mathbb{0}$ and $\mathbb{1}$ to avoid confusion with the number of Rydberg excitations. 
Figure~\ref{fig:1}(c) illustrates an example for a measured superatom syndrome. 
Superatoms that do not have exactly one Rydberg excitation violate the exactly-one constraint and are flagged with an error syndrome $\sigma_S = \mathbb{1}$, indicated in orange.

With this, the Rydberg-Parity embedding itself acts as an error-correcting code, the \textit{Rydberg-Parity code}. 
However, there is a major challenge to implementing the error correction: the code is \textit{nonlinear}. In particular, our error syndrome is based on number-checks instead of parity-checks in standard stabilizer codes. 
The allowed low-energy states of superatoms, which follow Eq.~\eqref{eq:constraint_superatom}, form the \textit{codewords}, but they do not form a linear subspace, unlike in linear codes.
For example, two exactly-one constraint-satisfying configurations of the unweighted dimer construct the codewords
\[
W = \{\ket{01}, \ket{10} \}\, ,
\]
which are not closed under bitwise addition,
\[
\ket{01 \oplus 10} = \ket{11} \notin W\, ,
\]
since $\ket{11}$ violates the exactly-one constraint and is, thus, not a codeword.
Consequently, we cannot apply available decoders designed for linear codes to correct errors, but instead design a tailored decoder for Rydberg gadget-based embeddings.

\subsection{\cluster decoder}

\begin{figure*}[t]
    \centering
    \includegraphics[width=1\linewidth]{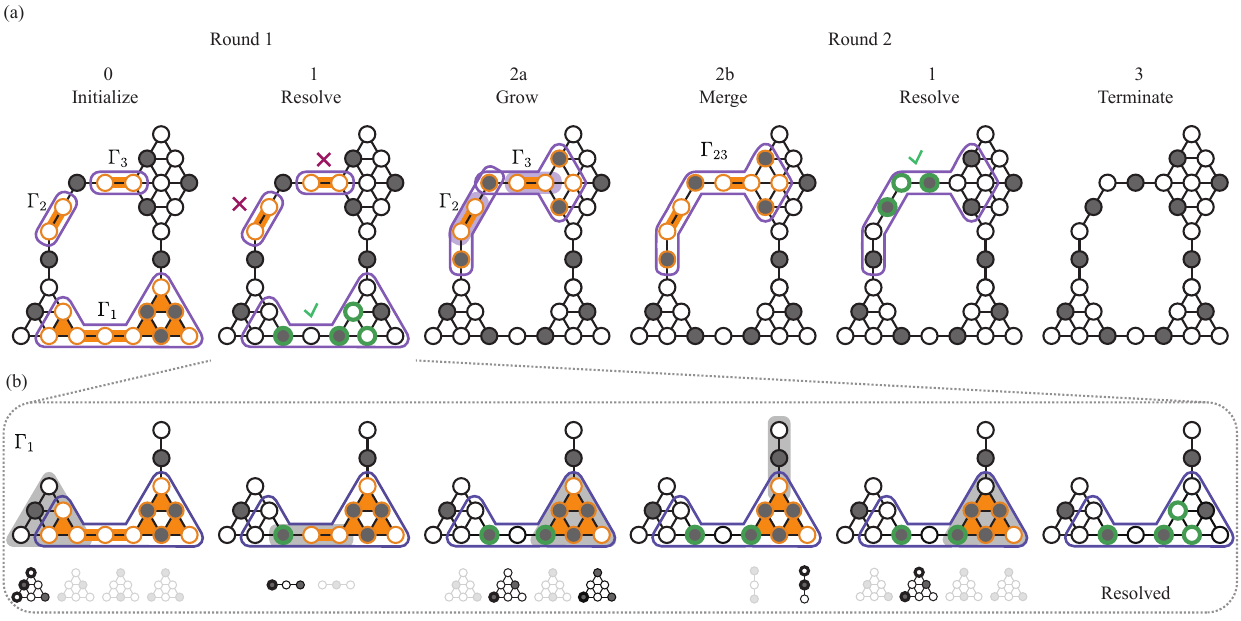}
    \caption{
    \cluster decoder. 
    (a) The iterative decoding algorithm consists of four stages: 
    (0) Initialize: Initial clusters $\Gamma_i$ (the active regions of interest) are defined by grouping adjacent superatom sets (sets of atoms belonging to the same superatom) that are flagged by the syndrome (highlighted in orange). 
    (1) Resolve: The decoder attempts to locally correct the errors in each cluster by finding a unique bitstring configuration using the lookup-table search detailed in (b). All atoms outside the cluster are frozen, whereas the atoms inside the cluster remain unfrozen and are outlined in orange. If the Resolve step fails (e.g., $\Gamma_2$ and $\Gamma_3$), i.e., no unique valid configuration is found, the cluster remains unresolved and is passed to step (2).
    (2) Grow and Merge: Reshape unresolved clusters. (2a) Grow: include neighboring superatom sets in each cluster; (2b) Merge: merge overlapping clusters $\Gamma_2$ and $\Gamma_3$ into a single cluster $\Gamma_{23}$.
    If there are unresolved clusters, the next round of the loop starts from step (1) using updated clusters.
    (3) Terminate: The \cluster decoder terminates when either all clusters are resolved, returning a valid MWIS bitstring, or an unresolved cluster covers the entire graph, indicating decoding failure.
    (b) Lookup-table search for cluster $\Gamma_1$: The decoder checks all gadgets that overlap with the cluster. First, the left 3BODY gadget is evaluated: among the four valid isolated MWIS configurations, only one is consistent with the three frozen atoms outside the cluster (bottom, highlighted with thick outlines). This unique configuration is selected, and all atoms in the gadget are then frozen. The resulting boundary condition uniquely constrains the adjacent LINK gadget.
    For the right 3BODY gadget, however, no unique candidate can be identified. The decoder therefore postpones this gadget and checks the remaining neighboring gadgets, in this case, the right LINK. After resolving the LINK, the two frozen ports uniquely determine the MWIS configuration of the right 3BODY gadget. Finally, cluster $\Gamma_1$ is successfully resolved.
    }
    \label{fig:3}
\end{figure*}

Here, we introduce the \cluster decoder, specifically tailored for decoding the nonlinear superatom error syndromes in Rydberg gadget embeddings. 
The \cluster decoder operates through a sequence of iterative steps, as shown in the following scheme:

\begin{center}
\resizebox{\columnwidth}{!}{%
\begin{tikzpicture}[
    >=stealth,
    node distance=9mm and 7mm,
    box/.style={
        draw,
        rounded corners=7pt,
        align=center,
        minimum width=2.25cm,
        minimum height=0.95cm,
        font=\footnotesize,
        line width=0.8pt
    },
    arrow/.style={->, line width=1.0pt}
]

\node[box] (init) {0. Initialize};
\node[box, right=of init] (resolve) {1. Resolve};
\node[box, right=of resolve] (terminate) {3. Terminate};

\node[box, below=7mm of resolve] (grow) {
    2a. Grow\\
    2b. Merge
};

\draw[arrow] ([yshift=1mm]init.east) -- ([yshift=1mm]resolve.west);
\draw[arrow] ([yshift=1mm]resolve.east) -- ([yshift=1mm]terminate.west);

\draw[arrow]
    ([yshift=2mm]resolve.south east)
    .. controls +(7mm,0mm) and +(7mm,0mm) ..
    (grow.east);

\draw[arrow]
    (grow.west)
    .. controls +(-7mm,0mm) and +(-7mm,0mm) ..
    ([yshift=2mm]resolve.south west);

\end{tikzpicture}%
}
\end{center}
The decoder begins with an \textit{Initialize} step that forms initial \textit{clusters} around the detected superatom syndromes. 
It then enters an iterative loop comprising two main stages. First, the \textit{Resolve} step attempts to find a unique logical configuration for each cluster independently. Next, the \textit{Grow and Merge} step expands the boundaries of any unresolved clusters and fuses those that overlap as a result of this growth. 
Finally, the \textit{Terminate} step ends the cycle when either all clusters are successfully resolved, or an unresolved cluster spans the entire graph.
Figure~\ref{fig:3} illustrates this iterative process for a representative bit-flip error pattern, demonstrating successful termination after two rounds. 
The precise mechanics of each step are detailed in the following:

\paragraph*{0. Initialize:} 
At the start, we define the set of clusters $\mathbb{\Gamma}=\{\Gamma_1, \Gamma_2, \dots\}$ from the superatom syndrome obtained from Eq.~\eqref{eq:superatom_syndrome} for a given input bitstring $\mathbf{n}$ and a graph $G_{\text{Parity}}^{\text{Rydberg}}$.
A cluster $\Gamma_i$ is defined as the union of adjacent unsatisfied superatom sets,
\[
\Gamma_i = \bigcup_{S_j \in \mathcal{D}_i} S_j.
\]
Here, $\mathcal{D} = \{S_i : \sigma_{S_i} = \mathbb{1}\}$ is the set of unsatisfied superatom sets, and the $\mathcal{D}_i$ are its maximal connected components, such that $\mathcal{D} = \bigsqcup_i \mathcal{D}_i$, where $\bigsqcup$ denotes the disjoint union. 
Two superatom sets $S$ and $S'$ are considered adjacent when $S \cap S' \neq \emptyset$.
Figure~\ref{fig:3}(a) [step 0] illustrates the Initialize process for a specific error syndrome giving rise to three clusters, $\mathbb{\Gamma} = \{\Gamma_1, \Gamma_2, \Gamma_3\}$.

\paragraph*{1. Resolve:} 
For each cluster $\Gamma_i$, we check if a \textit{unique} configuration that satisfies all its internal superatom constraints exists, without violating any external ones.
To this end, we \textit{freeze} all atoms outside the clusters, meaning that their states are fixed and assumed to not contain bit-flip errors during this step [see Fig.~\ref{fig:3}(b), where frozen (unfrozen) atoms are outlined in black (orange)].
Using the states of these frozen atoms together with predefined lookup tables of valid configurations for each gadget type [in this case Eqs.~\eqref{eq:link_state} and \eqref{eq:3body_states}], the decoder determines whether a unique error-correcting configuration exists within the cluster and, if so, applies it immediately.

In more detail, for a given cluster, we start from an arbitrary gadget that contains both frozen and unfrozen atoms, i.e., atoms outside and inside that cluster, respectively. The states of the frozen atoms then restrict the possible configurations of that gadget.
If only a single configuration is consistent with the frozen bit values, it is selected, and all remaining atoms in the gadget are assigned their corresponding bit values and frozen.
If no configuration is consistent, this means that no valid solution exists, and the Resolve step is aborted for this cluster.
If multiple configurations remain possible, the decoder leaves this gadget unresolved and proceeds to neighboring gadgets that have not yet been checked. After attempting to resolve these neighbors, it returns to the original gadget, where the updated information may further restrict its admissible configurations.
Unless the Resolve step has been aborted, the decoder then selects another gadget intersecting the cluster that contains both frozen and unfrozen atoms, and repeats the same procedure.

This step has three possible outcomes:
In the ideal case, a unique configuration is determined for all gadgets, yielding a single bitstring for the cluster;
at least one gadget may admit no valid configuration;
or, all unresolved gadgets admit multiple possible configurations.
In the last two cases, the cluster cannot be resolved and passes to the next step \textit{Grow and Merge}.

An example for the Resolve step is illustrated in Fig.~\ref{fig:3}(b) for cluster $\Gamma_1$. This cluster overlaps with four gadgets: the left 3BODY gadget, the middle LINK, the right 3BODY gadget, and the LINK attached at the top atom of the right 3BODY gadget. In this example, the decoder starts from the left 3BODY gadget. Since three atoms outside the cluster are frozen (highlighted with black and bold outlines), only a single one of the four valid configurations (bottom) matches the frozen atoms, and therefore can be uniquely chosen for the gadget. That freezes all atoms in this gadget for the subsequent steps.
After that, the middle LINK can be uniquely resolved in the same manner. 
However, the right 3BODY gadget cannot yet be uniquely resolved afterwards: with only one frozen port, two configurations are possible (bottom). The decoder therefore postpones the resolution of this gadget and checks the remaining one, the top LINK. 
That one can again be uniquely resolved and freezes the top port of the yet unresolved 3BODY gadget. This additional frozen port now uniquely determines its configuration, so that all gadgets have been uniquely resolved. Therefore, cluster $\Gamma_1$ is successfully resolved and removed from the cluster set $\mathbb{\Gamma}$. 

The clusters $\Gamma_2$ and $\Gamma_3$ cannot be resolved in this step because the frozen atoms in the LINK gadget do not impose a unique solution. These clusters are passed on to the next step.

Intuitively, whether a cluster resolves depends on where the errors lie relative to its boundary. 
During the Resolve step, all external atoms are frozen and considered as error-free.
If all errors lie strictly inside the cluster, the correct boundary conditions propagate inside the cluster yielding a single candidate solution. This is the case for $\Gamma_1$ in Fig.~\ref{fig:3}(b).
By contrast, if an error lies outside the cluster (such as between $\Gamma_2$ and $\Gamma_3$ in Fig.~\ref{fig:3}(a)) the assumption that the frozen boundary is error free is violated, preventing resolution during the Resolve step.
Only by enlarging clusters can such errors be brought inside them for later resolution.
Therefore, we \textit{Grow and Merge} the unresolved clusters in the next step.

\paragraph*{2. Grow and Merge:} 
This step enlarges the clusters so that the errors preventing their resolution are moved inside the clusters. It consists of two substeps:
\textit{(a) Grow}: Each cluster that was not resolved in the previous step expands by including its neighboring superatom sets,
\[
\Gamma_i \leftarrow \Gamma_i \cup
\bigcup_{\substack{S_j \nsubseteq \Gamma_i,\;  S_j \cap \Gamma_i \neq \emptyset}} S_j ,
\]
so that errors previously on the boundary could now lie in the bulk. 
\textit{(b) Merge}: Clusters enlarged in this way may overlap and, if so, are merged into a single larger one. The resulting set of clusters is used for the next round. 
An example is shown in Fig.~\ref{fig:3}(a) [steps 2a and 2b], where the clusters $\Gamma_2$ and $\Gamma_3$ that could not be resolved in the previous step are first grown and then merged into a single cluster, $\Gamma_{23}=\Gamma_2\cup\Gamma_3$, because the grown clusters overlap.

\paragraph*{3. Terminate:}
There are two conditions for terminating the iteration process, both checked at the start of the Resolve step.
The first condition, $\mathbb{\Gamma}=\emptyset$, implies that all clusters have been resolved and the decoding was successful, i.e., a valid logical state (a codeword) was found.
The second condition is that a single cluster covers the entire graph, $|\Gamma_i|=|G_{\text{Parity}}^{\text{Rydberg}}|$, meaning that the decoding failed to resolve all clusters. In this case, no frozen boundary remains to select a unique configuration, and decoding terminates as a failure.

Importantly, successful termination of the \cluster decoder does not guarantee that the original target codeword has been recovered. 
The decoder may successfully resolve the syndrome but converge to a different codeword. 
When this occurs, the decoder has not reverted the physical errors that actually happened, but instead introduced additional bit-flips to reach another valid codeword, thereby committing a logical error. 
Consequently, a decoding run yields one of three outcomes: (1) successful recovery of the target codeword; (2) convergence to an incorrect codeword (a logical error); or (3) failure to find any codeword. The performance of the \cluster decoder is measured by the frequency of the first outcome.

The time complexity of the \cluster decoder algorithm can be estimated as the product of the number of rounds it takes to converge and the runtime cost per round.
The cost per round is dominated by the \textit{Resolve} step, specifically by checking all the gadgets in a cluster for valid configurations.
Since lookup tables for the two gadget types (LINK and 3BODY) are constructed only once at the start, querying these tables incurs a constant cost of $\mathcal{O}(1)$. 
During propagation, a gadget is fixed only when the current boundary conditions leave a unique candidate; otherwise, it is deferred and revisited when additional boundary information becomes available from neighboring gadgets.
Because each gadget has only a constant number of ports, it can receive new boundary information only a constant number of times within a round, such that each gadget is queried at most $\mathcal{O}(1)$ times.
The number of gadgets in the embedding scales as $\mathcal{O}(N^2)$, imposing a total cost per round of $\mathcal{O}(N^2)$.
Importantly, the Resolve step propagates only unique assignments and never branches over multiple candidates, avoiding exponential search.
Independent clusters can additionally be processed in parallel.
To reach the worst-case number of rounds, a cluster must grow to cover the entire embedded graph. For the two-dimensional lattice-like Rydberg-Parity layouts considered here, the number of physical atoms scales with the area, $N'=\mathcal{O}(N^2)$, while the graph radius scales with the linear extent, $\mathcal{O}(\sqrt{N'})=\mathcal{O}(N)$, analogous to the area-radius scaling of a Euclidean disk.
Since each \textit{Grow} step expands a cluster by one neighboring superatom layer, at most $\mathcal{O}(N)$ rounds are required for a cluster to span the full layout.
Consequently, the total worst-case time cost of the \cluster decoder scales as
\[
\underbrace{\mathcal{O}(N)}_{\text{number of rounds}} \times \underbrace{\mathcal{O}(N^2)}_{\text{cost per round}} = \mathcal{O}(N^3).
\]

\subsection{Error correction threshold}
\label{sec:threshold}

\begin{figure}
    \centering
\includegraphics[width=1\linewidth]{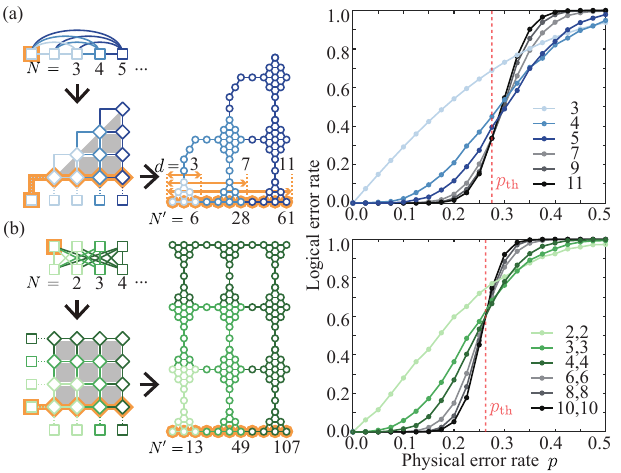}
    \caption{Logical error rate (LER) under i.i.d.~bit-flip noise for OPs on (a) complete graphs $G=K_N$ and (b) bipartite graphs $G=K_{N,N}$ for varying problem sizes $N$ (indicated by color). 
    The crossing of the curves reveals an error threshold $p_\text{th}$, below which the LER decreases monotonically as the system size $N$ increases.
    From the crossing of the curves corresponding to the two largest considered graph sizes, we estimate $p_{\text{th}} \approx 0.275$ for complete graphs (a), and $p_{\text {th}} \approx 0.263$ for bipartite graphs (b).
    The left side of each panel illustrates the mapping from the OP, through the parity-encoded problem, to the Rydberg-Parity embedding. Increasing problem sizes $N$ are represented by progressively darker colors.
    The orange highlights trace a single logical bit of the OP through its corresponding parity string in the parity-encoded problem to the associated chain of physical atoms in the Rydberg layout. 
    The code distance $d$ of the resulting Rydberg-Parity codes corresponds to flipping this indicated chain of atoms and is identical for $G=K_{N+1}$ and $G=K_{N,N}$.
    $N'$ denotes the physical size of the Rydberg-Parity codes.
    }
    \label{fig:4}
\end{figure}

To quantify the performance of the Rydberg-Parity code in combination with the specifically tailored \cluster decoder, we perform a numerical analysis of the logical error rate (LER).
Adapting standard error correction terminology, we refer to the original optimization problem on graph $G$ as the \textit{logical system}, while the embedded layout $G_{\text{Parity}}^{\text{Rydberg}}$ (the Rydberg-Parity code) is referred to as the \textit{physical system}. 
The specific properties of this code, especially the exact qubit overhead, the encoding rate, and the code distance, are strictly governed by the topology of $G$; therefore, we benchmark specific graph families.
We select two fundamental types for our analysis: complete graphs $G=K_N$ and bipartite graphs $G=K_{N,N}$, both of which require an embedding overhead of $\mathcal{O}(N^2)$.
To construct the physical Rydberg-Parity layouts here, we fix the basic LINK length to the minimum value $l_{\text{LINK}}=3$~\cite{Lanthaler2023}~\footnote{Note that the basic LINK length $l_{\text{LINK}}$ is the minimum length a LINK can have for horizontal and vertical connections between other gadgets. However, longer LINKs are still allowed in the physical layout to connect more distant gadgets, as seen in the Rydberg-Parity code for OPs on the complete graph $G=K_N$ [Fig.~\ref{fig:4}(a)].}.

Since the Rydberg-Parity code is nonlinear, the distance to the nearest competing codeword, $d(\mathbf{n})=\min_{\mathbf{n}'\neq\mathbf{n}} d_{\rm H}(\mathbf{n},\mathbf{n}')$, depends on the specific reference codeword $\mathbf{n}$.
Here, $d_{\rm H}(\mathbf{n},\mathbf{n}')$ denotes the Hamming distance between two bitstrings $\mathbf{n}$ and $\mathbf{n'}$. 
The global code distance $d$ is defined by minimizing $d= \min_\mathbf{n} d(\mathbf{n})$ over all codewords.
For $K_{N+1}$ and $K_{N,N}$, this yields a code distance scaling linearly with the size of the logical problem:
\begin{eqnarray}
  d = (1+l_{\text{LINK}})N - 2l_{\text{LINK}} + 1.
  \label{Eq:codedist}
\end{eqnarray}
This distance corresponds to an error vector spanning the atomic chain along the bottom boundary of the Rydberg-Parity codes, as shown in Fig.~\ref{fig:4}. 
For the chosen $l_{\text{LINK}}=3$, we obtain $d=4N-5$. 
Further details regarding the code distance of $G_{\text{Parity}}^{\text{Rydberg}}$ are provided in App.~\ref{App:codedist}.

To benchmark the \cluster decoder, we compare the LER against the physical bit-flip error rate $p$. 
We model the noise as independent and identically distributed (i.i.d.) bit-flip errors applied to each physical atom with probability $p$. 
For simplicity, we assume symmetric errors, $p(0|1) = p(1|0) = p$, where $p(0|1)$ and $p(1|0)$ denote the false-negative and false-positive probabilities, respectively.
The LER is estimated as the average frequency with which a logical error occurs after running the \cluster decoder on randomly sampled physical error patterns over $n_{\text{samp}}$ samples.
Specifically, for each independent trial, we randomly select a valid logical bitstring (codeword) $\mathbf{n}$, apply i.i.d.~bit-flip errors with probability $p$ per physical site, run the \cluster decoder, and record a success if and only if the decoder strictly returns the original codeword $\mathbf{n}$. 
If the decoder converges to a different codeword $\mathbf{n}' \neq \mathbf{n}$ or fails to converge entirely, it is counted as a logical error. 
We simulate physical error rates from $p = 0$ to $p = 0.5$ in steps of $0.025$, averaging over $n_{\text{samp}} = 5000$ independent trials for each value of $p$.

Figure~\ref{fig:4} shows the extracted LER as a function of the physical error rate $p$ for varying system sizes $N$. 
Both graph topologies exhibit a clear error-correction threshold behavior.
Specifically, the lines for different $N$ cross at a critical threshold error rate $p_{\text{th}}$.
Below this threshold, the LER decreases monotonically as the system size $N$ increases, demonstrating that the Rydberg-Parity code together with the \cluster decoder achieve scalable error correction.
For the complete graph $K_N$ [Fig.~\ref{fig:4}(a)], the critical threshold emerges at $p_{\text{th}} \approx 0.275$, estimated by the crossing of the curves for $N=9$ and $N=11$. 
Similarly, the bipartite graph $K_{N,N}$ [Fig.~\ref{fig:4}(b)] exhibits a threshold at a comparable $p_{\text{th}} \approx 0.263$, estimated by the crossing of the curves for the largest considered graphs with $N=8$ and $N=10$. 
Notably, the smallest systems in both cases ($K_3$ and $K_{2,2}$) show significantly higher LERs than their larger counterparts. 
This is a finite-size boundary effect: the minimal $K_3$ corresponds to a single 3BODY gadget. If the first round of decoding fails, the cluster instantly expands to encompass the entire system; without frozen atoms to constrain the search space, all MWIS candidates become equally valid, significantly reducing the probability of recovering the exact initial state. 
A similar finite-size boundary issue affects $K_{2,2}$.

Furthermore, we observe that the LER for bipartite graphs ($G=K_{N,N}$) is generally larger than for complete graphs ($G=K_N$). 
We attribute this effect to the higher encoding rate of the Rydberg-Parity code for bipartite graphs. 
The denser encoding leads to a higher multiplicity of uncorrectable low-weight error patterns [cf.~App.~\ref{App:codedist}]. 
Thus, even though the minimal code distance $d$ is identical, the tighter packing of codewords in the Rydberg-Parity code for bipartite OPs results in a larger LER.
Importantly, this effect does not alter the qualitative error-suppression scaling, i.e., the monotonic error reduction with increasing $N$ below threshold.

The error-correction capability of the code can be tuned further.
Longer LINKs naturally increase redundancy, and Eq.~\eqref{Eq:codedist} shows that the code distance indeed scales linearly with the LINK length, $d \propto l_{\text{LINK}}$. 
Interestingly, our numerical simulations reveal that as $l_{\text{LINK}}$ increases, the critical threshold $p_{\text{th}}$ also increases, a highly desirable feature that goes beyond simple distance scaling [see App.~\ref{App:linklen}].
Furthermore, the threshold behavior observed here is not unique to Rydberg-Parity  embeddings: the \cluster decoder achieves similar threshold behavior on other blockade-based Rydberg gadget layouts, such as crossing lattices and 3D quantum wire embeddings (see App.~\ref{App:CRS} and \ref{App:Wire}).

In summary, these numerical results highlight a major strength of our co-design approach. By actively exploiting the atom redundancy of the Rydberg embedding as an error-correcting code, the specifically tailored \cluster decoder achieves favorable error correction scaling with a remarkably high threshold of $p_{\text{th}} \gtrsim 26\%$.
This demonstrates that Rydberg gadget embeddings in neutral atom platforms can support scalable, robust quantum optimization through classical post-processing.

\section{QPU data results} \label{Sec:QPU}

\begin{figure*}[t]
    \centering
    \includegraphics[width=1\linewidth]{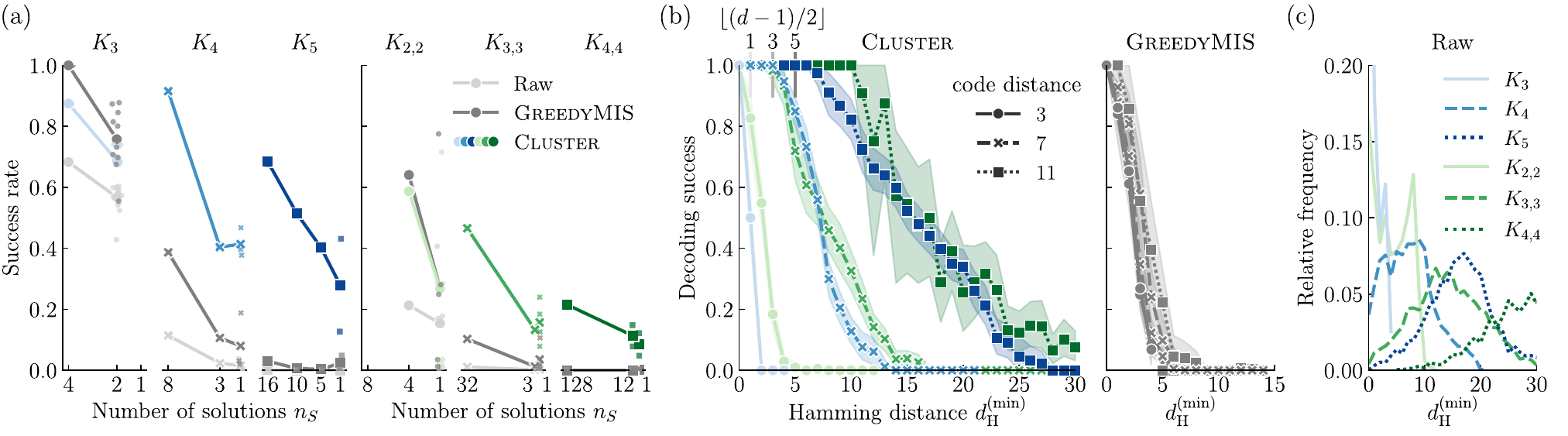}
    \caption{Benchmarking the Rydberg-Parity code with the \cluster decoder on experimental data from adiabatic quantum optimization on a QuEra Aquila QPU~\cite{Wurtz2023Aquila}.
    (a) Success rate of finding the target states for various problem instances on complete graphs $G=K_N$ (left) and bipartite graphs $G=K_{N,N}$ (right) of different sizes $N$. Individual problem instances are represented by small dots, plotted as a function of the number of solutions $n_S$ for that given instance. Large dots and solid lines show the average results for a given $n_S$. Raw experimental data are shown in light gray, data post-processed by the \ebadi algorithm are shown in dark gray, and \cluster decoded data are shown in colors corresponding to their problem class [cf.~Fig.~\ref{fig:4}].
    (b) Average success probability to obtain the target states after \cluster decoding (left) and \ebadi post-processing (right), plotted as a function of the minimal Hamming distance $d_{\rm H}^{({\rm min})}$ between the raw experimentally measured bitstring and the target state(s). Data are averaged over all problem instances within each problem class. Error bands indicate 95\% confidence intervals.
    (c) Relative frequency of measuring raw bitstrings at a given minimal Hamming distance from the target bitstring(s), averaged over all problem instances within each problem class. The distributions are normalized to sum up to unity.
    }
    \label{fig:experiment}
\end{figure*}

The existence of a finite threshold under the symmetric i.i.d.~error model shown in Sec.~\ref{sec:threshold} demonstrates that the Rydberg-Parity code with the proposed \cluster decoder supports scalable error correction. 
However, in real neutral-atom-based QPUs, various error sources invalidate this simple error model~\cite{DeLeseleuc2018, Tsai2025}. 
SPAM errors are inherently asymmetric due to atom loss and decay, such that $p(0|1) \neq p(1|0)$.
Optical crosstalk arising from spatial beam imperfections violates the identically distributed assumption.
Significantly, the independence assumption is broken by two primary sources of correlated errors: domain walls generated by diabatic dynamics through the Kibble-Zurek mechanism~\cite{Bernien2017, Keesling2019, Bombieri2025}, and phase-flip ($Z$) errors appearing during the adiabatic sweep that can ultimately map to correlated bit-flips at readout.

We now evaluate our error-corrected approach to quantum optimization using the \textit{QuEra Aquila} QPU~\cite{Wurtz2023Aquila}, leveraging its local detuning control $\Delta_i(t)$ to directly implement Rydberg gadgets  [see App.~\ref{App:ExpSet} for details on the experimental setting].
As in Sec.~\ref{sec:threshold}, we embed QUBO problems on complete ($K_N$) and bipartite ($K_{N, N}$) graphs via the Rydberg-Parity embedding~\cite{Lanthaler2023}. 
For thorough benchmarking, we select multiple problem instances with coefficients $J_i \in \{-1, 0, 1\}$ [summarized in Tab.~\ref{tab:instances} in App.~\ref{App:ExpSet}] that span different numbers of (degenerate) solutions, degrees of frustration in the parity-mapped problems, and requirements for local addressing beams.
For each instance layout, we execute a standard three-step, piecewise linear sweep, terminating at the target local detunings $\Delta_i$ to correctly embed the OP. 
Over $n_{\rm shots}$ repetitions, the final atom configurations are measured, yielding raw experimental bitstrings. 
We subsequently post-process these bitstrings using the proposed \cluster decoder and the standard \ebadi algorithm~\cite{Ebadi2022, Wurtz2024}, detailed in App.~\ref{app:ebadi}, to compare their error correction performance.

Figure~\ref{fig:experiment} shows a summary of our results for complete graphs $G=K_3, \, K_4, \, K_5$ with atom counts $N'=6, \, 28, \, 61$, and bipartite graphs $G=K_{2,2}, \, K_{3,3}, \, K_{4,4}$ with atom counts $N'=13, \, 49, \, 107$, respectively.
In Fig.~\ref{fig:experiment}(a) we plot the success rate of obtaining the target atom bitstrings as a function of the number of solutions $n_S$ of the implemented OP instance (small dots).
Specifically, we compare the success rate of the raw experimental data (light gray) and \ebadi post-processed data (dark gray) versus \cluster decoded results (colored).
Large symbols with connecting lines show the average success rate over all problem instances for the given graph and $n_S$.
We observe a strong decay of the raw experimental success rate with increasing system size and with decreasing $n_S$.
This is expected because we use fixed, finite sweep durations of $T=4\, \mu\rm{s}$ throughout all these experiments. 
Therefore, the annealing sweep generally becomes less adiabatic as the system size increases, due to decreasing energy gaps during the dynamics.
As expected, both post-processing methods improve the success rate relative to the raw experimental data.
The success rate after applying the \ebadi algorithm still decays strongly with system size, reaching zero or nearly zero for the largest layouts, $K_5$ and $K_{4,4}$, with 61 and 107 atoms, respectively. In contrast, the success rate after applying the \cluster decoder decreases more slowly, remaining at several percent even in the worst case.
Only for the smallest considered graphs, $K_3$ and $K_{2, 2}$, can \ebadi post-processing compete with the \cluster decoder. 
Importantly, the gap between raw or \ebadi post-processed results and \cluster-decoded results widens with the size of the logical system: the larger the problem, the more the decoder gains. 
This demonstrates the algorithmic scalability of the decoder, indicating that the theoretical benefits identified in Fig.~\ref{fig:4} survive real-world QPU noise.

This advantageous scalability is mainly rooted in the error-correcting properties of the atomic layouts as Rydberg-Parity codes.
For the implemented layouts, the code distance $d$ increases with $N$ in both graph families [Eq.~\eqref{Eq:codedist}]. 
Specifically, $d=3$ for $K_3$ and $K_{2,2}$, $d=7$ for $K_4$ and $K_{3,3}$, and $d=11$ for $K_5$ and $K_{4,4}$. 
The increasing code distance allows the \cluster decoder to successfully correct error patterns with larger Hamming weights.
Figure~\ref{fig:experiment}(b) shows the probability of recovering a target bitstring after \cluster decoding (left) or \ebadi post-processing (right), averaged over all problem instances for each graph. We plot this probability against the minimum Hamming distance $d_{\rm H}^{({\rm min})}$ between the raw experimental bitstring and the solution bitstring(s).
We observe that the \cluster decoder can successfully correct all bitstrings up to a given Hamming distance $d_{\rm H}^{({\rm min})}$ which, importantly, increases with the code distance $d$. 
This stands in stark contrast to the \ebadi post-processing [right panel in Fig.~\ref{fig:experiment}(b)], where the average success rate drops to zero already at $d_{\rm H}^{({\rm min})} \approx 5$ independent of $d$.
These results demonstrate that only the \cluster decoder efficiently exploits the underlying error-correcting code, directly driving the significantly enhanced success rates for the OPs shown in Fig.~\ref{fig:experiment}(a).

For a given code distance $d$ [cf.~Eq.~\eqref{Eq:codedist}] one can theoretically correct up to $\lfloor(d-1)/2\rfloor$ errors [indicated by short vertical lines at the top of Fig.~\ref{fig:experiment}(b)].  
Interestingly, for the largest considered problem instances $K_5$ and $K_{4,4}$ with $d=11$, the \cluster decoder successfully corrects all measured bitstrings up to $d_{\rm H}^{({\rm min})}>5$, exceeding the guaranteed correction range.
We also observe similar behavior in numerical simulations with i.i.d. noise, as shown in App.~\ref{App:codedist}. 
Therefore, we relate this observation to the features of the error vectors, i.e., patterns of bit-flip errors that transform
one valid codeword into another one, in the nonlinear Rydberg-Parity codes. 
In particular, only a few error vectors have Hamming weight $d$, while many others have larger Hamming weights [cf.~Fig.~\ref{fig:code_distance}(b) in App.~\ref{App:codedist}].
Thus, many error patterns can still be corrected even if they have more than $\lfloor(d-1)/2\rfloor$ errors.

In Fig.~\ref{fig:experiment}(c) we plot the relative frequency of obtaining raw experimental bitstrings at a given $d_{\rm H}^{({\rm min})}$ for the different implemented graphs. 
We clearly observe that the histograms are skewed to increasingly larger $d_{\rm H}^{({\rm min})}$ with increasing system size. 
We attribute this effect to larger diabatic errors due to the limited annealing duration, combined with increasing experimental noise as the number of atoms grows.
The improved performance of the \cluster decoder with increasing $d$ cannot fully compensate for the additional errors accumulated as system size grows. Consequently, the overall success rate decreases with system size, as observed in Fig.~\ref{fig:experiment}(a). 
This indicates that the overall effective bit-flip error rate, arising from diabatic errors and experimental imperfections, is currently beyond threshold for these experiments. 
However, several caveats to this argument must be considered. 
First, the effective bit-flip errors at the end of the annealing sweep are not purely i.i.d., unlike the assumption used to derive the theoretical threshold in Sec.~\ref{sec:threshold}. 
Second, the effective error rate per physical qubit might not remain constant as the system size increases, due to the highly nonlinear mapping of physical errors to logical bit-flips at the end of the sweep. 
Finally, applying the \cluster decoder remains highly beneficial even in regimes where effective error rates exceed the threshold.
In standard error correction approaches, one would avoid the qubit overhead from using a code by directly working on bare physical bits if the error rate is beyond threshold.
However, in our case, the Rydberg-Parity embedding is strictly mandatory just to map the original optimization problem onto the Rydberg atom QPU. 
Therefore, this spatial overhead cannot be avoided, and applying the decoder comes at no extra hardware cost and strictly improves the raw experimental results.

In summary, these experimental results demonstrate that the \cluster decoder can efficiently exploit the error-correcting structure of Rydberg-Parity embeddings under realistic QPU noise to improve quantum optimization success rates.
For the largest layout of 107 atoms, it recovers target solutions with an average success rate of $\gtrsim8\%$, even when the raw and \ebadi-processed data find none.
This confirms that the error-correcting properties identified in Sec.~\ref{sec:error_correction} remain practically useful on present-day neutral-atom hardware.
Finally, let us note that the \cluster decoder was recently also applied to the AQC results obtained in a globally driven, Parity Architecture-based neutral-atom array in Ref.~\cite{Lanthaler2026}, yielding a significant improvement in the target-state probability.

\section{Conclusions and Outlook}
\label{sec:conclusion}

We have presented an error correction scheme for quantum optimization on Rydberg atom arrays with gadget-based embeddings, leveraging the native Rydberg blockade to detect and correct errors. 
We have shown that the gadget embedding itself functions as a nonlinear classical error-correcting code, providing error syndromes directly from locally counting Rydberg excitations. To exploit these syndromes, we have developed a purpose-built decoder, the \cluster decoder, that efficiently corrects bit-flip errors.
This approach represents a genuine hardware-software co-design: the physical layout of the Rydberg atoms acts as the hardware-level MWIS solver, while the embedded spatial structure simultaneously serves as the error-correcting code.
We have shown that the \cluster decoder exhibits a finite error threshold on the Rydberg-Parity codes of $p_{\text{th}} \gtrsim 26\%$ under an i.i.d. bit-flip noise model.
Furthermore, we have demonstrated the effectiveness of this approach for quantum optimization on the QuEra Aquila device, showing significant improvements over standard post-processing methods (in particular the \ebadi algorithm) in terms of overall success rate.
This performance gain stems directly from the fact that the \cluster decoder recognizes and utilizes the underlying error-correcting code structure of the Rydberg layouts.

In the main text, we have focused on applying the \cluster decoder to Rydberg-Parity layouts. 
More generally, however, the \cluster decoder applies to layouts built from a finite set of bounded-size gadgets whose admissible configurations are known and propagate deterministically from one gadget to the next. 
For example, the crossing lattice scheme~\cite{Nguyen2023, DeOliveira2025, Bombieri2025} satisfies both conditions, while the quantum wire approach~\cite{Qiu2020, Kim2022, Byun2022} to quantum optimization satisfies only the first one, resulting in an additional cost for a lookup table that grows linearly with the length of the wire.
The application of the \cluster decoder to these alternative approaches is analyzed in App.~\ref{App:CRS} and App.~\ref{App:Wire}.
For unstructured Rydberg layouts, these conditions generally do not hold. 
While blockade violations remain detectable, a satisfied syndrome no longer strictly certifies a codeword. 
Extending the decoder to such unstructured layouts remains an open task for future work.

Our methodology draws a parallel to graph reduction techniques~\cite{Schuetz2025, Schuetz2025qReduMIS}, which similarly exploit the Rydberg blockade constraint that only one atom within a superatom can belong to the MWIS. 
However, while reduction techniques use this constraint to simplify the graph, our approach uses it to suppress errors. In this way, we treat the auxiliary atoms introduced by the embedding as a resource for error correction rather than as an overhead. 
Whether the two philosophies can be combined~\cite{kombe2026}, and if so, how the resulting structure trades array size against logical protection, remains open. 
Quantifying this trade-off on present-day QPUs is a natural next step.

Furthermore, our setting can be contrasted with quantum annealing correction~\cite{Jordan2006, Ganti2014} under minor embedding (QAC-ME)~\cite{Pudenz2014, Pudenz2015, Vinci2015, Vinci2016, MunozBauza2025}. 
There, minor embedding is applied first to realize the required connectivity, and a second layer of encoding is added on top via finite ferromagnetic couplings that impose energy penalties for broken stabilizers.
The corresponding decoding algorithm is typically energy minimization (EM)~\cite{Vinci2015}, which restricts the decoding problem to qubits flagged as broken and solves it via simulated annealing.
These energy penalties create a soft constraint where a disagreement is penalized in energy rather than strictly forbidden.
Our exactly-one excitation condition on superatoms, in contrast, admits no physical violation. 
The two decoders thus sit at different points of the same trade-off: EM always returns an answer but provides no strict guarantees, whereas every cluster resolved by our decoder is mathematically guaranteed to be a codeword of the embedded graph.

Finally, we note that we utilized typical three-step, piecewise-linear pulses for the adiabatic evolution.
It is well known that optimized pulse shaping can significantly improve AQC outputs~\cite{Finzgar2024, Leclerc2025, Perseguers2025, Senoo2026, Zhang2026}. 
Exploring the interplay, or potential trade-offs, between optimized analog pulse shapes and digital decoding success probabilities presents a highly promising avenue for future research.

\begin{acknowledgements}
We thank Evgeny Ostroumov for initial contributions to this project.
We thank Martin Lanthaler, Kilian Ender, Paul Schnabl, Christophe Goeller, Jonathan Wurtz, and JuYoung Park for fruitful discussions.
The authors gratefully acknowledge their colleagues at ParityQC for their technical contributions and software development efforts that supported this work.
During the preparation of this work, the authors used generative AI tools to improve the wording and clarity of the manuscript and to assist with software development. The authors reviewed and edited the resulting content and take full responsibility for the final publication.
This research was funded in whole, or in part, by the Austrian Science Fund (FWF) SFB BeyondC Project No. F7108-N38 (DOI: 10.55776/F71). 
This project was supported by FFG Fundings (Project Nos. FO99918691 and FO999933929) as part of the international Eureka cooperation. 
This study was supported by the Austrian Research Promotion Agency (FFG Project No. FO999937388, FFG Basisprogramm).
This research is funded by the German Federal Ministry of Research, Technology and Space (BMFTR) within the project MUNIQC-ATOMS (Project No. 13N16080).
P.I.\ has received funding from the European Union (Horizon-MSCA-Doctoral Networks) through the project QLUSTER (HORIZON-MSCA-2021-DN-01-GA101072964).
\end{acknowledgements}

\bibliography{bibliography.bib}

\appendix

\section{Embedding strategies} \label{App:embedding}

An \textit{embedding}~\cite{Choi2008} is a map between graphs:
\[
\Phi_G : G \rightarrow G'.
\]
Given an optimization problem $\mathrm{OP}$ on $G$ and $\mathrm{OP}'$ on $G'=\Phi_G(G)$, the embedding $\Phi_G$ induces a corresponding map on the solution space,
\[
\Phi_N : \mathcal{N}_G \rightarrow \mathcal{N}_{G'},
\]
called the \textit{embedding map}~\cite{DeCorrec2025}, where $\mathcal{N}_G$ and $\mathcal{N}_{G'}$ are the sets of feasible bitstrings of $\mathrm{OP}$ and $ \mathrm{OP}'$, respectively. We require $\Phi_N$ to satisfy
\[
\mathbf{n}_1 \ne \mathbf{n}_2 \implies \Phi_N(\mathbf{n}_1) \ne \Phi_N(\mathbf{n}_2),
\]
for all $\mathbf{n}_1, \mathbf{n}_2 \in \mathcal{N}_G$, and to preserve the ordering of the cost functions $f_{\text{cost}}^{\text{OP}} : \mathcal{N}_G \to \mathbb{R}$ and $f_{\text{cost}}^{\text{OP}'} : \mathcal{N}_{G'} \to \mathbb{R}$ of $\mathrm{OP}$ and $\mathrm{OP}'$, respectively, i.e.,
\[
f_{\text{cost}}^{\text{OP}}(\mathbf{n}_1) \le f_{\text{cost}}^{\text{OP}}(\mathbf{n}_2)
\implies
f_{\text{cost}}^{\text{OP}'}(\Phi_N(\mathbf{n}_1)) \le f_{\text{cost}}^{\text{OP}'}(\Phi_N(\mathbf{n}_2)),
\]
for all $\mathbf{n}_1, \mathbf{n}_2 \in \mathcal{N}_G$. In addition, every optimal solution of $\mathrm{OP}'$ must either belong to the image of $\Phi_N$ or be logically equivalent to a configuration in its image due to gadget degeneracy; no other configuration may attain the optimum of $\mathrm{OP}'$. In the present context, the target graph $G'$ is chosen to be compatible with the geometric constraints of a Rydberg atom array.

In the main text, we focus primarily on the Rydberg-Parity embedding. Here, we first review the mapping to the Parity Architecture that precedes its Rydberg atom implementation, and then consider two alternative Rydberg blockade-based embedding strategies: the crossing lattice and 3D quantum wire. For the latter two cases, we focus on how their local gadget structures allow the embedded layouts to be regarded as error-correcting codes and enable the application of the \cluster decoder.

\subsection{Parity Architecture} \label{App:Parity}

The Rydberg-Parity embedding considered in the main text is constructed in two steps [Eq.~\eqref{eq:parity_embedding}]. Here, we briefly review the first step,
\[
    G \rightarrow G_{\text{Parity}}.
\]

Consider an optimization problem represented by a pairwise Ising Hamiltonian on a graph $G=(V,E)$,
\begin{equation}
    \hat{H}_{\text{Ising}}
    =
    \sum_{(i,j)\in E} J_{ij}\hat{Z}_i\hat{Z}_j .\nonumber
\end{equation}
The central idea of the Parity Architecture is to encode each interaction term into a parity qubit. For a two-body interaction, this mapping is
\begin{equation}
    \tilde{Z}_{(i,j)} \leftarrow \hat{Z}_i\hat{Z}_j .\nonumber
\end{equation}
Accordingly, the interaction terms of the original problem become local fields acting on the corresponding parity qubits [Fig.~\ref{fig:1}(a)].

The resulting parity qubits are not independent, but must satisfy parity constraints originating from the logical variables on $G$. For example, three parity qubits associated with a closed loop obey [Fig.~\ref{fig:1}(a)]
\begin{equation}
    \tilde{Z}_{(1,2)}
    \tilde{Z}_{(1,3)}
    \tilde{Z}_{(2,3)}
    =
    \tilde{\mathbb{1}} .\nonumber
    \label{eq:parity_closedloop}
\end{equation}
Such relations define the logical subspace of the Parity Architecture. They can be implemented as local three- and four-body parity constraints, yielding a Hamiltonian of the form
\begin{subequations}
\begin{eqnarray}
    \tilde{H}_{\text{Parity}}
    &=&
    \sum_l J_l \tilde{Z}_l
    +
    \tilde{H}_{\text{Const}} \nonumber \\
    \tilde{H}_{\text{Const}} &\propto & - \sum_{\text{loops}} \tilde{Z}_{l_1} \tilde{Z}_{l_2} \tilde{Z}_{l_3} (\tilde{Z}_{l_4}), \nonumber
\end{eqnarray}
\end{subequations}
where $\tilde{H}_{\text{Const}}$ energetically enforces the local parity constraints. For an all-to-all connected problem, a standard local realization is provided by the Lechner-Hauke-Zoller (LHZ) architecture~\cite{Lechner2015}.

More generally, parity compilation~\cite{Ender2023,terHoeven2023} determines an appropriate set of parity qubits and local constraints for arbitrary interaction graphs while reducing unnecessary qubit overhead. 
For the bipartite graph considered in the main text, parity compilation is applied~\cite{Dreier2026} to generate a qubit-optimal layout.
The same construction also extends to higher-order terms by associating a parity qubit with the corresponding product of logical operators, e.g.,
\begin{equation}
    \hat{Z}_i\hat{Z}_j\hat{Z}_k
    \rightarrow
    \tilde{Z}_{(i,j,k)} . \nonumber
\end{equation}
This completes the first mapping step, $G\rightarrow G_{\text{Parity}}$; its Rydberg atom implementation is described in Sec.~\ref{sec:gadgets}.

\subsection{Crossing lattice} \label{App:CRS}

\begin{figure}
    \centering
\includegraphics[width=1.0\linewidth]{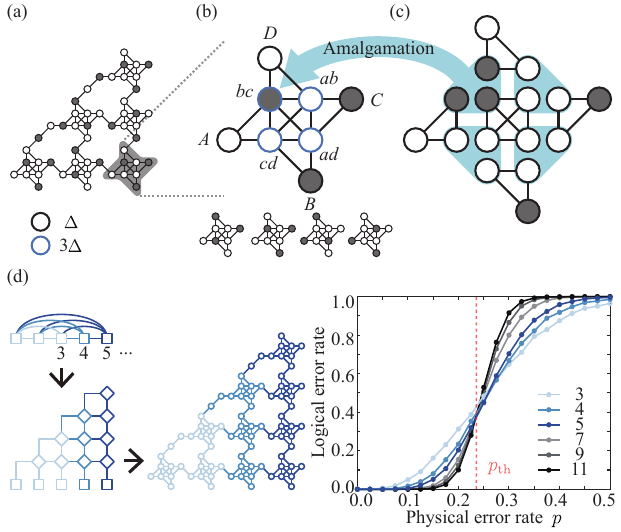}
    \caption{(a) Crossing lattice embedding. (b) CROSSING (CRS) gadget. (c) Superatom-amalgamation structure of the CRS gadget. (d) Threshold behavior with $p_{\text{th}}\approx0.235$. }
    \label{fig:CRS}
\end{figure}

Crossing lattice embedding realizes nonlocal connectivity in a planar Rydberg atom array using two types of gadgets: even-atom LINK and CROSSING (in short, CRS) gadgets [Fig.~\ref{fig:CRS}(a)]. LINK gadgets propagate logical information across the lattice, while CRS gadgets allow two logical lines to cross while preserving the corresponding logical relations.

A CRS gadget simultaneously implements NOT relations \footnote{In Ref.~\cite{Stastny2023}, this gadget is referred to as the inverted crossing (ICRS) gadget.} between the pairs $(n_A, n_C)$ and $(n_B, n_D)$ in a planar atomic array,
\[
C_{\text{CRS}}: (n_A = 1 - n_C) \wedge (n_B = 1 - n_D). 
\]
It is composed of four logical bits $(n_A, n_B, n_C, n_D)$ with detuning $\Delta$, and four auxiliary atoms $(n_{ab}, n_{bc}, n_{cd}, n_{ad})$ with detuning $3\Delta$ [Fig.~\ref{fig:CRS}(b)]. 
The blockaded Hamiltonian of the CRS gadget is
\begin{eqnarray}
\hat{H}_{\text{CRS}}^{\text{BLC}} &=& -\Delta (\hat{n}_A+\hat{n}_B+\hat{n}_C+\hat{n}_D) \nonumber \\
    & &-3\Delta (\hat{n}_{ab}+\hat{n}_{bc}+\hat{n}_{cd}+\hat{n}_{ad}) . \nonumber
\end{eqnarray}
The CRS gadget admits four MWIS configurations: 
\begin{eqnarray}
\ket{n_An_B;n_Cn_D;n_{ab}n_{bc}n_{cd}n_{ad}}=\ket{00;11;0010}, \nonumber \\
 \ket{01;10;0100}, \ket{10;01;0001}, \text{and }\ket{11;00;1000}.
 \label{eq:CRSstates}
\end{eqnarray}

As shown in Fig.~\ref{fig:CRS}(c), a CRS gadget can be decomposed into an amalgamation of superatoms:
\begin{eqnarray}
C_{\text{CRS}}^{\text{Rydberg}}&=& C_S(n_A,n_{bc},n_{cd})\wedge C_S(n_B,n_{cd},n_{ad})\wedge \nonumber \\
& &
C_S(n_C,n_{ab},n_{ad}) \wedge C_S(n_D,n_{ab},n_{bc}) \wedge \nonumber \\
& &
C_S(n_{ab},n_{bc},n_{cd},n_{ad}), \nonumber
\end{eqnarray}
with corresponding Hamiltonian,
\begin{eqnarray}
\hat{H}^{\text{BLC}}_{\text{CRS}}&=&\underbrace{-\Delta(\hat{n}_A+\hat{n}_{bc}+\hat{n}_{cd})}_{C_S(n_A,n_{bc},n_{cd})} \underbrace{-\Delta(\hat{n}_B+\hat{n}_{cd}+\hat{n}_{ad})}_{C_S(n_B,n_{cd},n_{ad})} \nonumber \\
& &
\underbrace{-\Delta(\hat{n}_{C}+\hat{n}_{ab}+\hat{n}_{ad})}_{C_S(n_{C},n_{ab},n_{ad})} \underbrace{-\Delta(\hat{n}_{D}+\hat{n}_{ab}+\hat{n}_{bc})}_{C_S(n_{D},n_{ab},n_{bc})} \nonumber \\
& &
\underbrace{-\Delta(\hat{n}_{ab}+\hat{n}_{bc}+\hat{n}_{cd}+\hat{n}_{ad})}_{C_S(n_{ab},n_{bc},n_{cd},n_{ad})}. \nonumber
\end{eqnarray}
Thus, both LINK and CRS gadgets are entirely composed of superatoms satisfying the exactly-one excitation constraint. The same superatom number check and \cluster decoder introduced in the main text can therefore be directly applied to the crossing lattice embedding without modification.

The crossing-with-edge (CWE) gadget~\cite{Nguyen2023, Bombieri2025}, an extension of the CRS gadget, can likewise be decomposed into superatoms and therefore treated within the same decoding framework.

We next evaluate the error-correction performance of the crossing lattice embedding by analyzing the LER under the same i.i.d. bit-flip noise model used in the main text. We set the LINK length to $l_{\text{LINK}}=2$. To apply the \cluster decoder, we use the CRS-gadget lookup tables provided in Eq.~\eqref{eq:CRSstates}. The resulting LER curves exhibit threshold behavior similar to that of the Rydberg-Parity embedding, with a threshold at $p_{\text{th}}\approx0.235$ [Fig.~\ref{fig:CRS}(d)]. In summary, the crossing lattice embedding provides a direct extension of our error-correction framework beyond the Rydberg-Parity Architecture, while retaining the same superatom syndrome and \cluster decoder procedure.

\subsection{3D Quantum wire} \label{App:Wire}

\begin{figure}
    \centering
\includegraphics[width=1.0\linewidth]{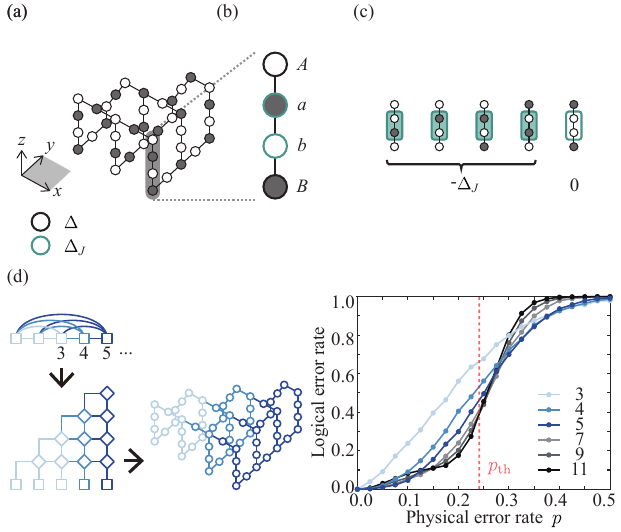}
    \caption{(a) Quantum wire embedding. (b) (Even-atom) WIRE gadget. (c) Energy assignment by the WIRE gadget. (d) Threshold behavior with $p_{\text{th}}\approx0.241$. }
    \label{fig:WIRE}
\end{figure}

A quantum wire is an atomic chain inserted between two distant nodes, which implements an effective interaction between them. By utilizing the three-dimensional configurability of a tweezer-arranged atomic array, quantum wire embeddings are not restricted to planar graphs and can access more general graph topologies. Quantum wires were originally proposed for a Rydberg-dressing scheme~\cite{Qiu2020} and later extended to Rydberg-blockade-based architectures~\cite{Kim2022, Byun2022}. Here, we focus on the blockade-based quantum wire.

To be consistent with the other embedding strategies considered in this work, we assume a lattice-like atomic layout with constant height, thereby retaining an overhead of $\mathcal{O}(N^2)$ [Fig.~\ref{fig:WIRE}(a)]. Quantum wire embedding requires two types of gadgets: LINK and WIRE. LINK gadgets form the lattice, while WIRE gadgets connect otherwise non-overlapping LINKs in three dimensions by utilizing the height direction. 
We assume a fixed WIRE length to retain the $\mathcal{O}(N^2)$ scaling of the total number of physical atoms.

There are two types of quantum wires~\cite{Byun2024, Byun2024HUBO}: even-atom and odd-atom quantum wires. For an even-atom quantum wire, consider the simplest case of length $l_{\text{WIRE}}=4$, where the ports $A$ and $B$ encode the corresponding bits of the original graph $G$ [Fig.~\ref{fig:WIRE}(b)]. As mentioned in the previous paragraph, this wire is amalgamated between two LINKs representing $A$ and $B$. The blockaded Hamiltonian is given by
\begin{eqnarray}
\hat{H}^{\text{BLC}}_{\text{WIRE}} \nonumber
=
-\Delta (\hat{n}_A+\hat{n}_B)
-\Delta_J (\hat{n}_a+\hat{n}_b).
\end{eqnarray}
For given port values $(n_A, n_B)$, the allowed configurations are the interior configurations that minimize $\hat{H}^{\text{BLC}}_{\text{WIRE}}$ [Fig.~\ref{fig:WIRE}(c)]:
\begin{equation}
\label{eq:wire_states}
\begin{aligned}
\ket{n_A;n_a n_b;n_B}
={}&\ket{0;01;0}, \ket{0;10;0}, \\
&\ket{0;10;1}, \ket{1;01;0},
\text{ and } \ket{1;00;1}.
\end{aligned}
\end{equation}

Note that the logical state $(n_A,n_B)=(0,0)$ admits two degenerate WIRE configurations, $\ket{0;01;0}$ and $\ket{0;10;0}$. Except for the configuration corresponding to $(n_A,n_B)=(1,1)$, all allowed configurations contain a single excitation within the WIRE and therefore acquire the same energy contribution $-\Delta_J$. In contrast, the $(1,1)$ configuration contains no excitation within the WIRE. Thus, up to this common energy offset and the local port terms, the wire effectively assigns the interaction
\[
\hat{H}^{\text{BLC}}_{\text{WIRE}}
\rightarrow
\Delta_J \hat{n}_A\hat{n}_B .
\]
By properly tuning $\Delta_J$, the quantum wire can therefore control the corresponding quadratic weight in a QUBO, providing not only connectivity but also tunability of the interaction strength. However, $\Delta_J$ is a detuning parameter that must satisfy the condition in Eq.~\eqref{eq:mwis_condition}; therefore, an arbitrary interaction weight cannot be assigned. To realize a quadratic interaction with the opposite sign, an odd-atom quantum wire is required. By the same argument, an odd-atom wire yields
\[
\hat{H}^{\text{BLC}}_{\text{WIRE}}
\rightarrow
-\Delta_J
(\hat{n}_A\hat{n}_B-\hat{n}_A-\hat{n}_B),
\]
which provides the opposite-sign quadratic term together with additional local terms on the two ports. The odd-atom wire admits a different set of valid configurations. For simplicity, we consider only the even-atom wire in our numerical analysis.

Unlike the superatom-based gadgets in the Rydberg-Parity and crossing lattice embeddings, a quantum wire does not generally satisfy an exactly-one excitation constraint because it admits kink configurations. Therefore, the superatom syndrome based on the exactly-one condition cannot be directly applied. Fortunately, a quantum wire permits only a finite set of configurations, allowing us to utilize this structural information. We therefore generalize the syndrome from an exact-number check to an allowed-configuration check.

For a WIRE gadget $W$, let $L_W$ denote the set of locally allowed configurations. For the $l_{\text{WIRE}}=4$ even-atom wire in Eq.~\eqref{eq:wire_states},
\begin{equation}
L_W =
\left\{
(0;01;0),
(0;10;0),
(0;10;1),
(1;01;0),
(1;00;1)
\right\}. \nonumber
\end{equation}
We then define the generalized WIRE syndrome as
\begin{equation}
\sigma_W(\mathbf{n}_W)=
\begin{cases}
\mathbb{0}, & \text{if } \mathbf{n}_W \in L_W,\\
\mathbb{1}, & \text{otherwise}. \nonumber
\end{cases}
\label{eq:wire_syndrome}
\end{equation}
Thus, $\sigma_W=\mathbb{0}$ if and only if the local WIRE configuration belongs to $L_W$.
In the Initialize step, a WIRE gadget with $\sigma_W=\mathbb{1}$ is treated as a violated local constraint and included in the corresponding error cluster. With this generalized local check, the \cluster decoder can be applied to quantum wire-based layouts in the same manner as in the main text. For WIRE gadgets only, the exactly-one number check is replaced by a check against the allowed configurations in $L_W$; all other superatom checks remain unchanged. During the Resolve step, only the WIRE configurations that are consistent with the boundary atom states are retained. Thus, only the local validity check and lookup table associated with WIRE gadgets are modified, while the overall \cluster decoder procedure remains unchanged.

Like the Rydberg-Parity and crossing lattice embeddings, the quantum wire embedding also exhibits threshold behavior after applying the \cluster decoder [Fig.~\ref{fig:WIRE}(d)]. We assume $l_{\text{LINK}}=3$ and $l_{\text{WIRE}}=4$, respectively. To quantify the LER, we choose a single canonical physical representative for each logical state. In particular, $\ket{0;01;0}$ and $\ket{0;10;0}$ represent the same logical state, $(n_A,n_B)=(0,0)$, and are both valid WIRE configurations. For the LER calculation, however, only one canonical representative is used for each logical state in both the initial sampling and decoding.  The curves exhibit a crossing at $p_{\text{th}}\approx0.241$. Unlike the other embeddings, however, we observe an additional crossing at $p\approx0.148$. A detailed understanding of this double-crossing behavior is left for future work.

\section{Code distance and error vectors of Rydberg-blockade-based Parity Architecture} 
\label{App:codedist}

\begin{figure}[t]
    \centering
    \includegraphics[width=1\linewidth]{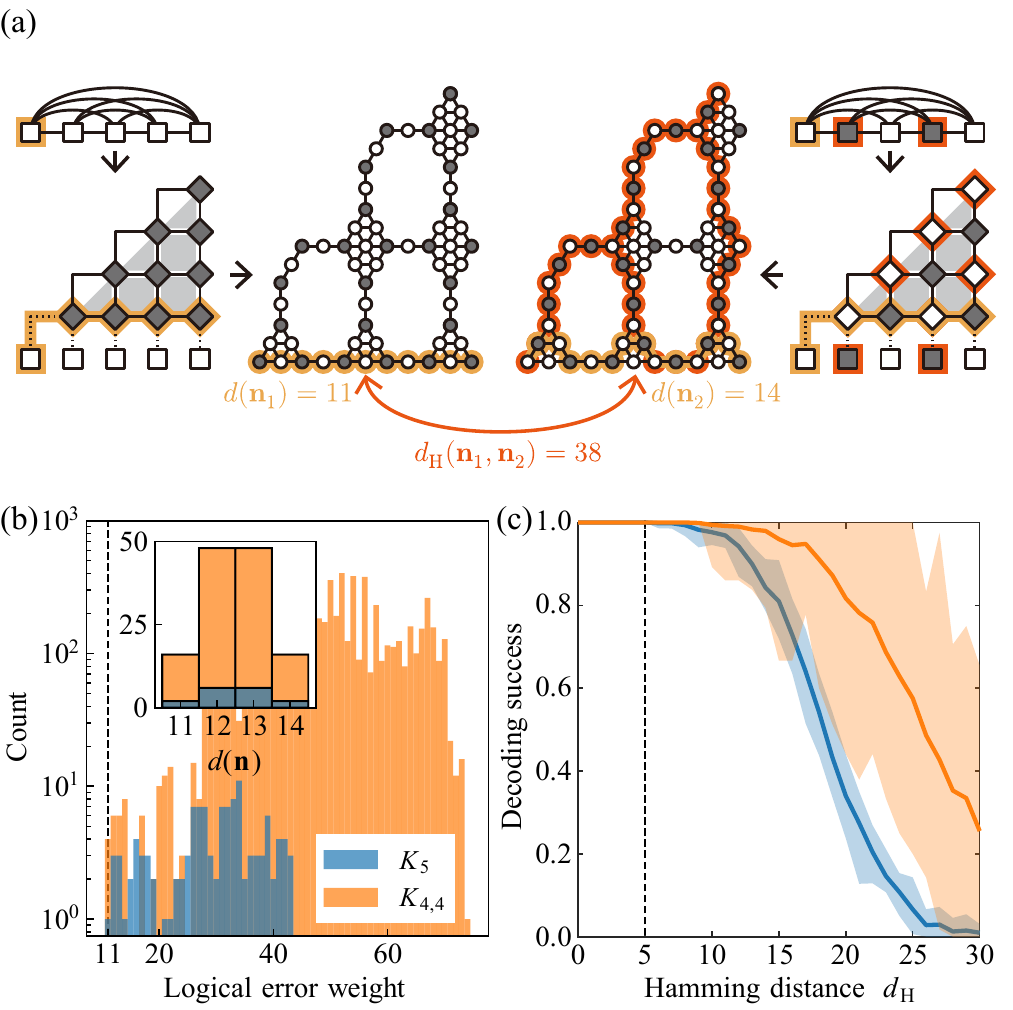}
    \caption{
    (a) Local distance and error vector. Exemplary codewords $\mathbf{n}_1$ and $\mathbf{n}_2$ for the logical system $G=K_5$: $\mathbf{n}_1$ has the shortest local distance, $d(\mathbf{n}_1)=d=11$, whereas $\mathbf{n}_2$ has the longest local distance, $d(\mathbf{n}_2)=14$. The error vector between these two codewords has Hamming weight $d_{\text{H}}(\mathbf{n}_1,\mathbf{n}_2)=38$, much larger than the code distance $d=11$.
    (b) Distribution of the Hamming weight of all error vectors between codewords in the Rydberg-Parity code for the all-to-all connected graph $K_5$ (blue) and the bipartite graph $K_{4,4}$ (orange). 
    The code distance $d=11$ for the problems is indicated by a vertical, dashed line.
    The inset shows the distribution of $d(\mathbf{n})$, i.e. the minimum error vector weight for each codeword.
    (c) \cluster decoder decoding success probability versus the Hamming weight $d_{\mathrm H}$ for the same $K_5$ and $K_{4,4}$ Rydberg-Parity codes. We use the $95\%$ percentile interval to characterize the variation in decoding success rates across logical states. All sampled errors are corrected beyond the conventional bound $\lfloor(d-1)/2\rfloor$ (dashed line, $d_{\mathrm H}=5$), consistent with the low multiplicity of minimum-weight error vectors shown in (b).
    }
    \label{fig:code_distance}
\end{figure}

As discussed in the main text, the inherent nonlinearity of the Rydberg-Parity code implies that the minimum number of bit-flips required to reach an adjacent codeword (i.e., a valid MWIS state) depends on the initial codeword.
For a given codeword $\mathbf{n}$, we define this local distance as $d(\mathbf{n})=\min_{\mathbf{n}'\neq\mathbf{n}} d_{\rm H}(\mathbf{n},\mathbf{n}')$, where $d_{\rm H}$ denotes the standard Hamming distance.
The global code distance $d$ characterizes the worst-case vulnerability of the embedding and is obtained by taking the minimum over all codewords, $d = \min_\mathbf{n} d(\mathbf{n})$.
For both the complete graph $K_{N+1}$ and the bipartite graph $K_{N,N}$, this global minimum scales linearly with the logical problem size according to
\[
  d = (1+l_{\text{LINK}})N - 2l_{\text{LINK}} + 1 \, .
\]

Figure~\ref{fig:code_distance}(a) illustrates two exemplary codewords, $\mathbf{n}_1$ and $\mathbf{n}_2$, for the logical system $G=K_5$.
The codeword $\mathbf{n}_1$ has the minimum local distance, $d(\mathbf{n}_1)=d=11$, whereas $\mathbf{n}_2$ has the maximum local distance, $d(\mathbf{n}_2)=14$.
Geometrically, the local distance is associated with an antiferromagnetically ordered line along the lower boundary of the layout.
This example clearly demonstrates the codeword dependence of the local distance.
Moreover, the Hamming distance between these two codewords is $d_{\mathrm H}(\mathbf{n}_1,\mathbf{n}_2)=38$, substantially larger than the code distance $d=11$.

The inset of Fig.~\ref{fig:code_distance}(b) illustrates the distribution of $d(\mathbf{n})$ for all codewords in the Rydberg-Parity codes corresponding to the complete graph $K_5$ and the bipartite graph $K_{4,4}$, both obeying a code distance of $d=11$ for the chosen link length $l_{\text{LINK}}=3$. 
In contrast to linear codes, $d(\mathbf{n})$ exhibits a broad distribution ranging from the code distance $\min(d(\mathbf{n})) = d = 11$ to $\max(d(\mathbf{n})) = 14$.
This demonstrates that many logical codewords possess an improved decodability compared to the worst-case vulnerability dictated by $d$. 

The main panel of Fig.~\ref{fig:code_distance}(b) shows the distribution of the Hamming weights for all \textit{error vectors} between every pair of codewords (i.e., the error patterns that transform one valid codeword into another) for the same Rydberg-Parity codes.
Although the minimum weight rigorously matches the code distance $d$, the distributions are remarkably broad, indicating that only a small fraction of error vectors have low Hamming weights.
While both problems have the same code distance $d$, the bipartite problem $K_{4,4}$ shows a significantly higher multiplicity of low-weight error vectors due to its higher encoding rate compared to $K_{5}$.
This directly increases the LER for a given physical error rate $p$, as observed in Fig.~\ref{fig:4} in the main text.
Importantly, this effect only shifts the curves but does not alter the qualitative emergence of a finite error threshold [cf.~Fig.~\ref{fig:4}, main text].

In Fig.~\ref{fig:code_distance}(c), we use the same set of Rydberg-Parity codes for $K_5$ and $K_{4,4}$ as in Fig.~\ref{fig:code_distance}(b) and analyze the success probability of the \cluster decoder as a function of the Hamming weight $d_{\rm H}$ between the noisy bitstring and the target codeword under i.i.d. bit-flip noise, with noisy bitstrings sampled uniformly over $d_{\rm H}$.
This numerical analysis is directly analogous to the experimental evaluation presented in Fig.~\ref{fig:experiment}(b) of the main text.
Similar to the experimental results, we observe that all sampled error patterns below a certain Hamming weight are successfully corrected by the \cluster decoder.
Interestingly, this empirical threshold extends beyond the expected value of  $\lfloor (d-1)/2 \rfloor$ correctable errors for a code with distance $d$, indicated by the dashed line at $d_{\rm H}=5$ in Fig.~\ref{fig:code_distance}(c).
We attribute this observation to the fact that error vectors of weight $d$ are rare within the full distribution of error vectors [cf.~Fig.~\ref{fig:code_distance}(b)] and are therefore rather unlikely to be sampled.

\section{LINK-length dependence of the threshold} \label{App:linklen}

\begin{figure}[t]
    \centering
\includegraphics[width=1.0\linewidth]{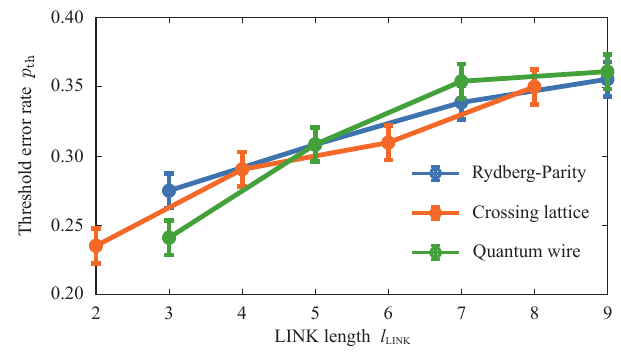}
    \caption{LINK-length dependence of the threshold. Error bars correspond to the sampling step size of the physical error rate.
    }
    \label{fig:Link_length}
\end{figure}

Interestingly, for all three embeddings considered here, the Rydberg-Parity, crossing lattice, and quantum wire embeddings, the threshold error rate $p_{\text{th}}$ increases monotonically with the LINK length $l_{\text{LINK}}$ after applying the \cluster decoder.

As shown in Fig.~\ref{fig:Link_length}, we test $l_{\text{LINK}}=3,5,7,9$ for the Rydberg-Parity and quantum wire embeddings, which require odd-length LINKs, and $l_{\text{LINK}}=2,4,6,8$ for the crossing lattice embedding, which requires even-length LINKs. For a fair comparison, we extract the threshold error rate $p_{\text{th}}$ from the crossing point between the results for the all-to-all connected graphs $G=K_N$ with $N=9$ and $N=11$, following the same procedure as in Figs.~\ref{fig:4}, \ref{fig:CRS}, and \ref{fig:WIRE}.

From the minimum to the maximum LINK length tested here, the threshold error rate increases from $p_{\text{th}}\approx0.275$ to $p_{\text{th}}\approx0.356$ for the Rydberg-Parity embedding, from $p_{\text{th}}\approx0.235$ to $p_{\text{th}}\approx0.350$ for the crossing lattice embedding, and from $p_{\text{th}}\approx0.241$ to $p_{\text{th}}\approx0.361$ for the quantum wire embedding.

\section{\ebadi algorithm}
\label{app:ebadi}

\ebadi is a heuristic, greedy algorithm-based correction method for M(W)IS problems, originally introduced in Refs.~\cite{Ebadi2022, Wurtz2024}. 
Here, we briefly summarize our implementation of that algorithm.
The algorithm is divided into two sequential steps: (1) greedy-removal and (2) greedy-addition.

First, given a noisy candidate bitstring, the greedy-removal process iteratively flips atoms that violate the Rydberg blockade condition from the excited state to the ground state, $\ket{1} \rightarrow \ket{0}$. 
Specifically, the algorithm counts the number of blockade violations for each atom and removes atoms in descending order of their violation count. 
In the case of a tie, one of the tied atoms is selected at random.
Following each atom flip, the blockade violations for the remaining atoms are efficiently updated.
After the greedy-removal step, no Rydberg blockade violations remain, and the bitstring corresponds to an independent set (IS).

Next, the greedy-addition process attempts to greedily add further atoms to the IS by flipping them from the ground state to the Rydberg state, $\ket{0} \rightarrow \ket{1}$, without introducing new blockade violations.
To do so, the algorithm first identifies a pool of candidate atoms that have no adjacent excited atom, ensuring they can be added to the IS without breaking the IS criterion.
Starting with the candidate atom that has the fewest adjacent candidates, we greedily add atoms to the set until the candidate pool is completely depleted.
In each step, the list of available candidates and their respective neighbor counts are dynamically updated.
At the end of this step a \textit{maximal} independent set (mIS; not to be confused with MIS) state is obtained by construction.

\section{Experimental setting} \label{App:ExpSet}

\begin{figure}[t]
    \centering
\includegraphics[width=0.7\linewidth]{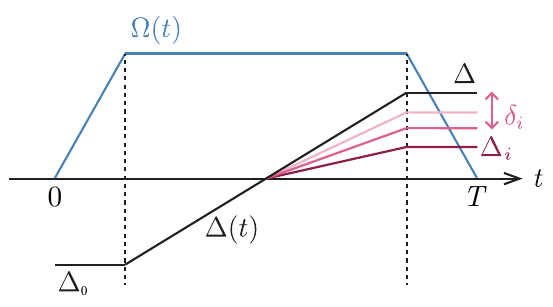}
    \caption{Schematic illustration of the state preparation pulse of duration $T$ used in this manuscript.
    The Rabi frequency $\Omega(t)$ follows a standard, three-step piecewise-linear protocol.
    The local detunings $\Delta_i(t)$ are composed of a global detuning component $\Delta(t)$ and local detuning shifts $\delta_i(t)\leq0$ which are turned on once the global detuning becomes positive.
    The final local detunings $\Delta_i(T)$ are set to match the target detunings $\Delta_i$ of the Rydberg-Parity embedding of the specific problem, where the maximum value is set as the final global detuning, $\max(\Delta_i) = \Delta = \Delta(T)$.
    }
    \label{fig:pulse}
\end{figure}

We perform all experiments presented in the main text on a QuEra Aquila QPU~\cite{Wurtz2023Aquila} featuring local detuning control.
To execute the quantum optimization protocol, we first arrange the atoms into the Rydberg-Parity embedded layout and initialize the atoms in their ground states $\ket{0\cdots0}$.
While the Rabi frequency $\Omega(t)$ can only be applied globally, the detuning can be locally tuned via a combination of a global drive $\Delta(t)$ and site-dependent local detuning shifts $\delta_i(t)$.
These local shifts are generated using spatially targeted light-shifting laser beams (via the AC Stark effect) and are restricted to negative offsets $\delta_i(t) \leq 0$.
The total instantaneous detuning on each atom $i$ is therefore given by $\Delta_i(t) = \Delta(t) + \delta_i(t)$.

Figure~\ref{fig:pulse} illustrates our state preparation schedule of total duration $T$.
We employ a standard three-step piecewise-linear profile for the Rabi frequency $\Omega(t)$, starting and ending at $\Omega(0) = \Omega(T) = 0$.
Initially, the local detuning shifts are turned off, $\delta_i(0)=0$, and the global detuning is set to $\Delta(0) = \Delta_0 < 0$, ensuring the initially prepared state $\ket{0\cdots0}$ is the instantaneous ground state of the system.
This detuning is held constant while the Rabi frequency ramps up to its maximum value to induce quantum fluctuations.
Subsequently, $\Delta(t)$ is linearly swept from $\Delta_0$ to a final value $\Delta(T) = \Delta > 0$.
Because the Rydberg-Parity embedding requires a specific pattern of target detunings $\Delta_i$ at the end of the sweep, and because $\delta_i(t) \leq 0$, we set the final global detuning to the maximum required target detuning: $\Delta = \max(\Delta_i)$.

To realize the smaller target detunings on the remaining atoms, we calculate the necessary relative shifts at the end of the pulse as $\delta_i = \Delta_i - \Delta$.
Dynamically, the local drives $\delta_i(t)$ are switched on only after the global detuning crosses zero, i.e., $\Delta(t) > 0$.
The time-dependent local shifts $\delta_i(t)$ are then scaled proportionally to the global detuning: $\delta_i(t) / \Delta(t) = \delta_i / \Delta$.
Once the detunings have reached their target values, they are kept constant while the Rabi frequency is slowly ramped down to zero.
Note that the local detunings are rapidly turned off over a period of $0.1\,\mu$s at the very end of the sweep (omitted from Fig.~\ref{fig:pulse} for visual clarity) to turn off the light-shifting lasers for subsequent measurements.
Small-scale numerical simulations (not shown) confirm that this rapid ramp-down does not introduce significant state dynamics, especially because of the short time period and the already very small Rabi frequency during this stage.

At the end of the protocol, a projective measurement is performed on the full array, yielding a raw bitstring of atom occupations $\mathbf{n} = \{n_i \in \{0, 1\}\ \}$.
By compiling statistics over independent experimental shots for each problem instance, the success rate is computed as the relative frequency of bitstrings $\mathbf{n}$ which correspond to one of the target MWIS bitstrings for the given problem.
For all presented results, we take $n_{\rm shots}=1000$ experimental runs. However, due to probabilistic atom loading, a fraction of these shots contain missing or improperly initialized atoms. We therefore post-select the data, retaining only those $n_{\rm shots}^{\rm full}$ shots where the entire array was successfully loaded prior to the adiabatic sweep [cf. Tab.~\ref{tab:instances}].
Additionally, we apply the \cluster decoder (or the \ebadi algorithm as a comparison) as a classical post-processing step on these raw bitstrings to extract the corrected configurations which are then used to evaluate the corresponding success probabilities.

Table~\ref{tab:instances} summarizes all experimental problem instances utilized in Sec.~\ref{Sec:QPU} in the main text.
Experimental benchmarking results for a single 3BODY gadget are presented below.

\begin{table*}[t]
\centering
\caption{Summary of problem instances for different graph layouts, showing the applied fields, the number of target solutions $n_S$, the number of broken constraints $f$ in the field-aligned configuration in the parity mapped problem, the total number of experimental shots $n_{\rm shots}$, and the number of experimental shots after post-selecting on initially fully filled arrays $n_{\rm shots}^{\rm full}$ which are used for subsequent analysis. We also report the peak global detuning magnitude $\Delta$ (the global detuning is swept from $-\Delta$ to $+\Delta$) and the peak local detuning shift $\delta_i^{\rm peak}$. All experiments use the interaction coefficient $C_6 = 2\pi \times 862690~\mathrm{MHz}\,\mu\mathrm{m}^6$, a peak Rabi frequency $\Omega_{\max} = 15.7$~rad/$\mu$s, and a nearest-neighbor interatomic distance of $6.4~\mu$m.}
\label{tab:instances}
\begin{tabular}{llcccccc}
\toprule
\textbf{Graph} & \textbf{Problem coefficients $\mathbf{J}$} & $\mathbf{n_S}$ & $\mathbf{f}$ & $\mathbf{n_{\rm shots}}$ & $\mathbf{n_{\rm shots}^{\rm full}}$ & $\boldsymbol{\Delta}$ (rad/$\mu$s) & $\boldsymbol{\delta_i^{\rm peak}}$ (rad/$\mu$s) \\
\midrule
$K_{3}$ & $(-x, 0, 0)$, $x \in \{0.1, 0.2, 0.4, 0.9, 0.98\}$ & 2 & 0 & 1000 & 735 - 773 & 54.6 & $-27.3$ \\
$K_{3}$ & $(0, 0, 0)$ & 4 & 0 & 1000 & 778 & 54.6 & $-27.3$ \\
$K_{3}$ & $(x, 0, 0)$, $x \in \{0.2, 0.4, 0.6, 0.8, 1.0\}$ & 2 & 0 & 1000 & 759 - 791 & 54.6 & $-27.3\,(1+x)$ \\
\midrule
$K_{4}$ & $(-1, -1, -1, 1, 1, -1)$ & 1 & 1 & 1000 & 607 & 40.0 & $-10.0$ \\
$K_{4}$ & $(-1, -1, 1, 1, 1, 1)$ & 1 & 1 & 1000 & 633 & 40.0 & $-40.0$ \\
$K_{4}$ & $(0, 0, 0, 0, 0, 0)$ & 8 & 0 & 1000 & 649 & 40.0 & $-20.0$ \\
$K_{4}$ & $(1, -1, -1, 1, 1, -1)$ & 1 & 2 & 1000 & 598 & 40.0 & $-40.0$ \\
$K_{4}$ & $(1, 1, 1, 1, 1, 1)$ & 3 & 2 & 1000 & 604 & 40.0 & $-40.0$ \\
\midrule
$K_{5}$ & $(-1, -1, -1, -1, -1, -1, -1, -1, -1, -1)$ & 1 & 0 & 1000 & 666 & 50.0 & $-20.0$ \\
$K_{5}$ & $(-1, -1, 1, 1, -1, 1, 1, 1, 1, 1)$ & 5 & 6 & 1000 & 681 & 50.0 & $-50.0$ \\
$K_{5}$ & $(-1, 1, -1, -1, 1, 1, 1, 1, 1, -1)$ & 1 & 4 & 1000 & 653 & 40.0 & $-40.0$ \\
$K_{5}$ & $(0, 0, 0, 0, 0, 0, 0, 0, 0, 0)$ & 16 & 0 & 1000 & 645 & 40.0 & $-20.0$ \\
$K_{5}$ & $(0, 0, 0, 0, 0, 0, 0, 0, 0, 0)$, [vdw compensated, see Ref.~\cite{Lanthaler2023}] & 16 & 0 & 1000 & 667 & 43.359 & $-23.359$ \\
$K_{5}$ & $(1, 1, 1, 1, 1, 1, 1, 1, 1, 1)$ & 10 & 3 & 1000 & 675 & 40.0 & $-40.0$ \\
\midrule
$K_{2,2}$ & $(-1, -1, -1, -1)$ & 1 & 0 & 1000 & 752 & 54.6 & $0.0$ \\
$K_{2,2}$ & $(-1, -1, -1, 1)$ & 4 & 1 & 1000 & 765 & 54.6 & $-54.6$ \\
$K_{2,2}$ & $(-1, 1, 1, -1)$ & 1 & 0 & 1000 & 822 & 54.6 & $-54.6$ \\
$K_{2,2}$ & $(1, -1, 1, -1)$ & 1 & 0 & 1000 & 779 & 54.6 & $-54.6$ \\
$K_{2,2}$ & $(1, 1, -1, -1)$ & 1 & 0 & 1000 & 789 & 54.6 & $-54.6$ \\
$K_{2,2}$ & $(1, 1, 1, 1)$ & 1 & 0 & 1000 & 816 & 54.6 & $-54.6$ \\
\midrule
$K_{3,3}$ & $(-1, -1, -1, 1, 1, 1, -1, -1, -1)$ & 1 & 0 & 1000 & 630 & 40.0 & $-10.0$ \\
$K_{3,3}$ & $(-1, -1, -1, 1, 1, 1, -1, 1, -1)$ & 1 & 2 & 1000 & 636 & 40.0 & $-40.0$ \\
$K_{3,3}$ & $(-1, 1, -1, 1, 1, 1, -1, 1, -1)$ & 1 & 4 & 1000 & 650 & 40.0 & $-40.0$ \\
$K_{3,3}$ & $(-1, 1, 1, 1, 1, 1, 1, 1, -1)$ & 3 & 2 & 1000 & 648 & 40.0 & $-40.0$ \\
$K_{3,3}$ & $(0, 0, 0, 0, 0, 0, 0, 0, 0)$ & 32 & 0 & 1000 & 653 & 40.0 & $-20.0$ \\
$K_{3,3}$ & $(1, 1, 1, 1, 1, 1, 1, 1, 1)$ & 1 & 0 & 1000 & 628 & 40.0 & $-40.0$ \\
\midrule
$K_{4,4}$ & $(-1, -1, -1, -1, -1, -1, -1, -1, -1, -1, -1, -1, -1, -1, -1, -1)$ & 1 & 0 & 1000 & 418 & 50.0 & $-20.0$ \\
$K_{4,4}$ & $(-1, 1, -1, 1, 1, 1, 1, 1, -1, 1, -1, 1, 1, 1, 1, 1)$ & 12 & 9 & 1000 & 410 & 50.0 & $-50.0$ \\
$K_{4,4}$ & $(0, 0, 0, 0, 0, 0, 0, 0, 0, 0, 0, 0, 0, 0, 0, 0)$ & 128 & 0 & 1000 & 446 & 40.0 & $-20.0$ \\
$K_{4,4}$ & $(1, -1, 1, -1, 1, 1, 1, 1, 1, 1, 1, 1, -1, 1, -1, 1)$ & 12 & 6 & 1000 & 432 & 40.0 & $-40.0$ \\
$K_{4,4}$ & $(1, 1, 1, 1, 1, 1, 1, 1, 1, 1, 1, 1, 1, 1, 1, 1)$ & 1 & 0 & 1000 & 418 & 40.0 & $-40.0$ \\
\bottomrule
\end{tabular}
\end{table*}

\subsection*{3BODY gadget benchmarking} \label{subsec:3body_experiment}

\begin{figure}[t]
    \centering
    \includegraphics[width=\columnwidth]{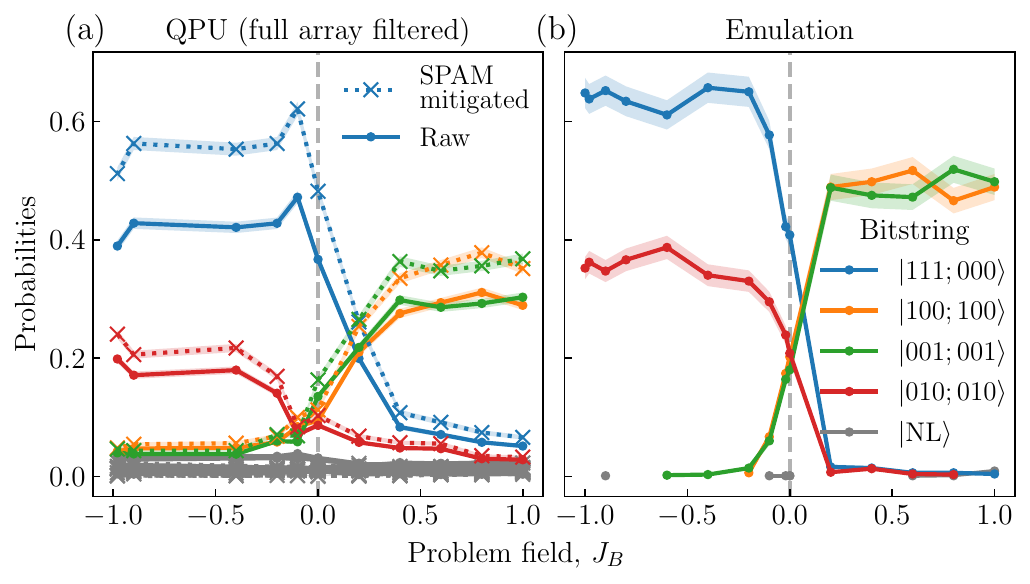}
    \caption{QPU (a) and emulation (b) results for the 3BODY gadget, where one \textit{problem field} $J_B$ (on atom B) is tuned. We show the probabilities of observing the four logical states and the six most likely non-logical states ($\ket{\rm NL}$) measured in the experiment versus the strength of the \textit{problem field}. For the QPU results, we show raw (SPAM mitigated) data with dots and full lines (crosses and dotted lines).
    }
    \label{fig:3body_experiment}
\end{figure}

We begin by benchmarking the 3BODY gadget consisting of six atoms on the QuEra Aquila QPU.
The 3BODY gadget is a central building block in the Rydberg-Parity embedding and implements a single parity constraint among three qubits, mapped to the edge atoms $A$, $B$, $AB$ [cf.~Fig.~\ref{fig:2}(c), main text]. The constraint is satisfied by four target states, which are the (nearly) degenerate ground states of the gadget when the blockade radius is appropriately set [Sec.~\ref{sec:gadgets}, main text]. For a sufficiently adiabatic state preparation protocol, these four states should be observed with high probability compared to higher-energy bitstrings.

Figure \ref{fig:3body_experiment}, at $J_B = 0.0$, shows the results of such an experiment with a pulse duration of $T=4\ \mu{\text s}$. 
We observe that the most likely bitstrings in the experiment [Fig.~\ref{fig:3body_experiment}(a)] and in the noise-free emulation results [Fig.~\ref{fig:3body_experiment}(b)] are the four target bitstrings $\left\{\ket{111;000},\ \ket{100;100},\ \ket{010;010},\ \ket{001;001} \right\}$, shown as colored lines.
Other bitstrings, shown in gray, are measured with significantly lower probabilities.
The probabilities of measuring the target bitstrings, however, are not uniform. These bitstrings follow different instantaneous energy curves during the sweep and become degenerate only at the end of the schedule.
So their dynamics differ, and the individual probabilities observed at the end of the sweep depend directly on the state preparation protocol. Fair sampling is, in general, hard to achieve in an adiabatic sweep~\cite {Sieberer2018, Konz2019}.

As detailed in App.~\ref{App:Parity}, the coefficients of a Rydberg-Parity embedded optimization problem are encoded as local fields $J_l$ on specific atoms $l$ of the embedded graph $G_{\text{Parity}}^{\text{Rydberg}}$, which energetically favor the bitstrings that solve the optimization problem.
As a first experimental benchmark, we vary a single field, $J_B\in[-1,1]$, acting on atom $B$ of the 3BODY gadget.
We use a fixed $T=4\,\mu\text{s}$, and compare the QPU results [Fig.~\ref{fig:3body_experiment}(a)] with noise-free emulations [Fig.~\ref{fig:3body_experiment}(b)].
As expected, a large negative local field $J_B < 0$ energetically favors bitstrings with an excitation on atom B, i.e.,  $\ket{111;000}$ (blue) and $\ket{010;010}$ (red). 
As we increase the field to positive values $J_B>0$, we observe a clear transition in the measured bitstring probabilities towards the target bitstrings of the optimization problem where atom $B$ is not excited, i.e., $\ket{100;100}$ (orange) and $\ket{001;001}$ (green).

QPU and emulation show a very good qualitative agreement.
Quantitatively, the experimentally measured target-bitstring probabilities are visibly reduced compared to the emulation due to noise.
Measurement errors, which occur after the end of the schedule, can be partially mitigated in post-processing when their rates are known~\cite{Nachman2020SPAM, Nation2021MTHREE}. 
Applying such mitigation with error rates $(p(1|0),p(0|1))=(0.01,\,0.08)$ (dotted lines and crosses in Fig.~\ref{fig:3body_experiment}(a)) increases the target-bitstring probabilities towards the emulation results.
The remaining discrepancy stems from noise that cannot be mitigated this way: Rydberg decay, laser noise (especially on the local detuning) and protocol imperfections, atomic motion, and atom placement errors.

These same experimental results are used to compute the raw, \ebadi post-processed, and \cluster decoded success rates for the graph $K_3$ in Fig.~\ref{fig:experiment} in the main text.

\end{document}